\documentclass[12pt]{article}
\usepackage{a4}
\usepackage{url,epsfig}
\usepackage{lineno}

\begin{document}
\title{Longitudinal phase-space matching \\
  between radio-frequency systems with different\\
  harmonic numbers and accelerating voltages}
\author{V. Ziemann, FREIA, Uppsala University}
\date{December 19, 2021}
\maketitle
\begin{abstract}\noindent
  We describe a simple mechanism to transform bunches with matched longitudinal
  phase-space distributions from one RF system to a matched distribution of a
  second RF system operating on a different harmonic and with a different
  accelerating voltage. The process is reversible and the longitudinal emittance
  is preserved. Simple equations that characterize the procedure are given and
  applications are discussed. In particular this method can be used to rapidly
  shorten the bunch length in proton rings.
\end{abstract}
%
%
\section{Introduction}
Handing over a beam from one RF system to a second one operating at a different frequency
and voltage frequently occurs when transferring beams from one ring to another or to
manipulate the bunch length in order to satisfy experimental requirements. Of course it is
desirable that the beam quality is preserved in these transitions, in particular, the
phase-space area occupied by the beam---the longitudinal emittance---should remain unchanged.
To analyze this scenario we consider a beam, controlled by one radio-frequency (RF) system,
operating with harmonic number $h_1$ and voltage $V_1$ to a second system, operating with
harmonic $h_2$ and voltage $V_2$. The dynamics of particles in a RF system---or in the
longitudinal phase space of a storage ring---is described by
\begin{equation}\label{eq:pendulum}
  \ddot\phi +\Omega_s^2\sin\phi=0
  \qquad\mathrm{with}\qquad
  \Omega_s^2 = -\frac{\omega_{\mathrm{rf}}\eta\cos\phi_s}{\beta^2T_0}\frac{e\hat V}{E_0}
\end{equation}
with the synchrotron frequency $f_s=\Omega_s/2\pi$, the RF frequency $f_{\mathrm{rf}}=\omega_{\mathrm{rf}}/2\pi$
and voltage $\hat V$, the phase slip factor $\eta$, the design phase $\phi_s$, the speed of the protons
$\beta=v/c$, the beam energy $E_0$, and the revolution time $T_0$. We assume that we operate 
under conditions where $\eta\cos\phi_s<0$ such that the stable phase is at $\phi=0$. We will
always use $\Omega_s$ at the first harmonic and voltage $\hat V$ as reference and note that
the synchrotron tune $\Omega(v,h)$ at voltage $v\hat V$ and harmonic $h$ is related to $\Omega_s$
by $\Omega(v,h)=\sqrt{vh}\Omega_s$. Henceforth we will always use the lower-case relative voltages
in this report.
\par
For the transitions we take inspiration from the 'bunch-muncher' described in~\cite{BM}, which
was used to shorten the bunches extracted from the damping rings of the SLAC linear collider
by briefly reducing the RF voltage and then turning it back on to rotate the bunch in longitudinal
phase space. To analyze this scenario and to derive equations that govern the transitions
we employ a linearized theory in the following section.
\section{Linear theory}
If we can assume that the bunches are short compared to the wavelength of the RF systems the
dynamics of the particles is governed by the linearized equation $\ddot\phi +\Omega_s^2\phi=0$
and their motion can be described by the transfer matrix
\begin{equation}
  \hat R(t)=\left(
    \begin{array}{cc} 
      \cos\left(\Omega_st\right) &  \sin\left(\Omega_st\right)/\Omega_s\\
      -\Omega_s\sin\left(\Omega_st\right) & \cos\left(\Omega_st\right)
    \end{array}
  \right)\ .                                      
\end{equation}
that operates on the state of $(\phi,d\phi/dt)$. If we consider motion at another harmonic
number $h$ and voltage $v$, we need to adapt the matrix to
\begin{eqnarray}\label{eq:TM}
  R(h,v,t) &=&
               \left(\begin{array}{cc} 1/h & 0 \\ 0 & 1 \end{array} \right)
  \left(
    \begin{array}{cc}
      \cos\left(\Omega t\right) &  \sin\left(\Omega t\right)/\Omega\\
      -\Omega\sin\left(\Omega t\right) & \cos\left(\Omega t\right)
    \end{array}
  \right)                                                      
  \left(\begin{array}{cc} h & 0 \\ 0 & 1 \end{array}\right)\nonumber\\
  &=&
  \left(
    \begin{array}{cc}
      \cos\left(\sqrt{vh}\Omega_st\right) &  \sin\left(\sqrt{vh}\Omega_st\right)/\sqrt{vh^3}\Omega_s\\
      -\sqrt{vh^3}\Omega_s\sin\left(\sqrt{vh}\Omega_st\right) & \cos\left(\sqrt{vh}\Omega_st\right)
    \end{array}
  \right)\ ,
\end{eqnarray}
where we use the abbreviation $\Omega=\sqrt{vh}\Omega_s$. The two outer matrices in the first
equality are necessary to scale the phase at harmonic $h$, because there are $h$ buckets in
the longitudinal extent where there is only one bucket at the first harmonic. Essentially
we stretch the phase-axis first, then we rotate, and finally we transform back into the
phase space of the first harmonic.
\par
It is now straightforward to verify that a matched beam $\sigma(v,h)$ is proportional to
\begin{equation}
\sigma(h,v)=\left(\begin{array}{cc}1/\sqrt{vh^3}\Omega_s & 0\\0&\sqrt{vh^3}\Omega_s\end{array}\right)\ .
\end{equation}
It reproduces after transport with $R(v,h,t)$ from Equation~\ref{eq:TM}, such that it obeys
$\sigma(h,v)=R\sigma(h,v)R^{\top}$.
\par
Our task is now to find a transfer matrix $M$ that transforms one matched beam with parameters
$h_1$ and $v_1$ to another one with $h_2$ and $v_2$, or $\sigma(h_2,v_2)=M\sigma(v_1,h_1)M^{\top}.$
In the 'bunch-muncher' from~\cite{BM} the voltage $v_1$ of the first RF system is lowered to
$\hat v$ for some time $t_1$ and later restored to its original value $v_1,$ which results in a
shortened bunch. If we now realize that choosing the time $t$ to obey $\sqrt{vh}\Omega_st=\pi/2$
the transfer matrix $R$ becomes
\begin{equation}\label{eq:S}
S(h,v)=\left(\begin{array}{cc}
    0 & 1/\sqrt{vh^3}\Omega_s \\ -\sqrt{vh^3}\Omega_s & 0
  \end{array}\right)\ .
\end{equation}
We now build our transforming system from two such matrices, one with a reduced voltage $\hat v$
and duration $t_1=\pi/\left(2\sqrt{\hat vh_1}\Omega_s\right)$ and a second one with restored
voltage $v_1$ and duration $t_2=\pi/\left(2\sqrt{v_1h_1}\Omega_s\right)$. The transfer matrix
$M$ for the transition, which we will call a {\em munch}, is then given by
\begin{equation}\label{eq:T}
  M=S(h_1,v_1)S(h_1,\hat v)
  =\left(\begin{array}{cc} -\sqrt{\hat v/v_1} & 0 \\ 0 & -\sqrt{v_1/\hat v}\end{array}\right)\ .
\end{equation}
The condition for $\sigma(h_2,v_2)=M\sigma(h_1,v_1)M^{\top}$ to be valid then leads to the
condition
\begin{equation}
  \sqrt{v_2h_2^3} \Omega_s = \frac{v_1}{\hat v}\sqrt{v_1h_1^3} \Omega_s
  \qquad\mathrm{or}\qquad
  \frac{\hat v}{v_1} =\sqrt{\frac{v_1}{v_2}} \left(\frac{h_1}{h_2}\right)^{3/2}\ ,
\end{equation}
which gives us the voltage during the munch $\hat v$ that is needed to transform the matched
longitudinal phase space of the first RF system to that of the second system. This condition
and the respective durations are summarized in Table~\ref{tab:munch}.
\par
\begin{table}[tb]
\begin{center}
  \begin{tabular}{l|ccl}
    State & $h$   & $v$    & Duration\\
    \hline
    Initially in first system  & $h_1$ & $v_1$  & arbitrary\\
    First part of transfer  & $h_1$
       & $\hat v=v_1\sqrt{\frac{v_1}{v_2}} \left(\frac{h_1}{h_2}\right)^{3/2}$
       & $t_1=\pi/\left(2\sqrt{\hat vh_1}\Omega_s\right)$\\
    Second part of transfer & $h_1$ & $v_1$ & $t_2=\pi/\left(2\sqrt{v_1h_1}\Omega_s\right)$\\
    Finally in second system & $h_2$ & $v_2$ & arbitrary
  \end{tabular}
\end{center}
\caption{\label{tab:munch}The parameters for a transition from on RF system to another one.}
\end{table}
In passing we note that the matrix $M$ is diagonal, but at the same time, is defined as
the product of two off-diagonal matrices from Equation~\ref{eq:S}. This resembles the
construction of the (transverse) transfer matrix for a telescope that is based on
two modules, each consisting of a thin lens with focal length $f$, sandwiched between
drift spaces of length $f$. Such a module has zeros on the diagonal and $f$ and $-1/f$
on the off-diagonal. Two such modules with focal length $f_1$ and $f_2$ then produce
the matrix for the telescope that has $-f_1/f_2$ and its inverse on the diagonal. We note
that in Equation~\ref{eq:S} the term $1/\sqrt{vh^3}\Omega_s$ takes the role of the focal
length $f$. The bunch-muncher can thus be understood as a telescope in longitudinal
phase space.
\par
The bunch length $\tilde \sigma_i=\sqrt{\sigma(h_i,v_i)}$ changes from one step to the next
by the magnification $|M_{11}|$ from Equation~\ref{eq:T}, such that we obtain
\begin{equation}\label{eq:demag}
  \frac{\tilde\sigma_2}{\tilde \sigma_1}=\sqrt{\frac{\hat v}{v_1}}
  =\left(\frac{v_1}{v_2}\right)^{1/4} \left(\frac{h_1}{h_2}\right)^{3/4}\ .
\end{equation}
We now need to see whether the linearized theory also works if we use the dynamics of the 
non-linear system of the mathematical pendulum from Equation~\ref{eq:pendulum}. 
\section{Model}
During the transitions only one RF system operates at a time, such that the dynamics of the
protons in longitudinal phase-space is governed by Equation~\ref{eq:pendulum}, which
describes the phase space of this mathematical pendulum and also synchrotron oscillations.
A characteristic feature of the dynamics is the existence of a separatrix, given for the
first harmonic by the equation
\begin{equation}
\dot\phi=\pm 2\Omega_s\cos(\phi/2)\ .
\end{equation}
This separatrix separates the periodic from the unbounded phase-space trajectories. Moreover,
closed-form solutions of the equations of motion for the pendulum equation, expressed in
terms of Jacobi-elliptic functions~\cite{ABRASTE}, are available~\cite{VZAPB}. Using these
solutions makes step-by-step integration obsolete, but only work if the design phase $\phi_s$
is zero and only if one RF system is operational at a time. These conditions apply in our
case and we therefore adapt the MATLAB~\cite{MATLAB} software that accompanies~\cite{VZAPB}
to handle different harmonics $h$ and voltages $v$.
%
\section{Simulations}
Here we consider several transitions from one RF system to another, where we always use the
bunch-munch equations from Table~\ref{tab:munch}, but use the closed-form solution of the
pendulum equation to follow a large number of sample particles. 
\subsection{Voltage step}
%
\begin{figure}[tb]
\begin{center}
\includegraphics[width=0.47\textwidth]{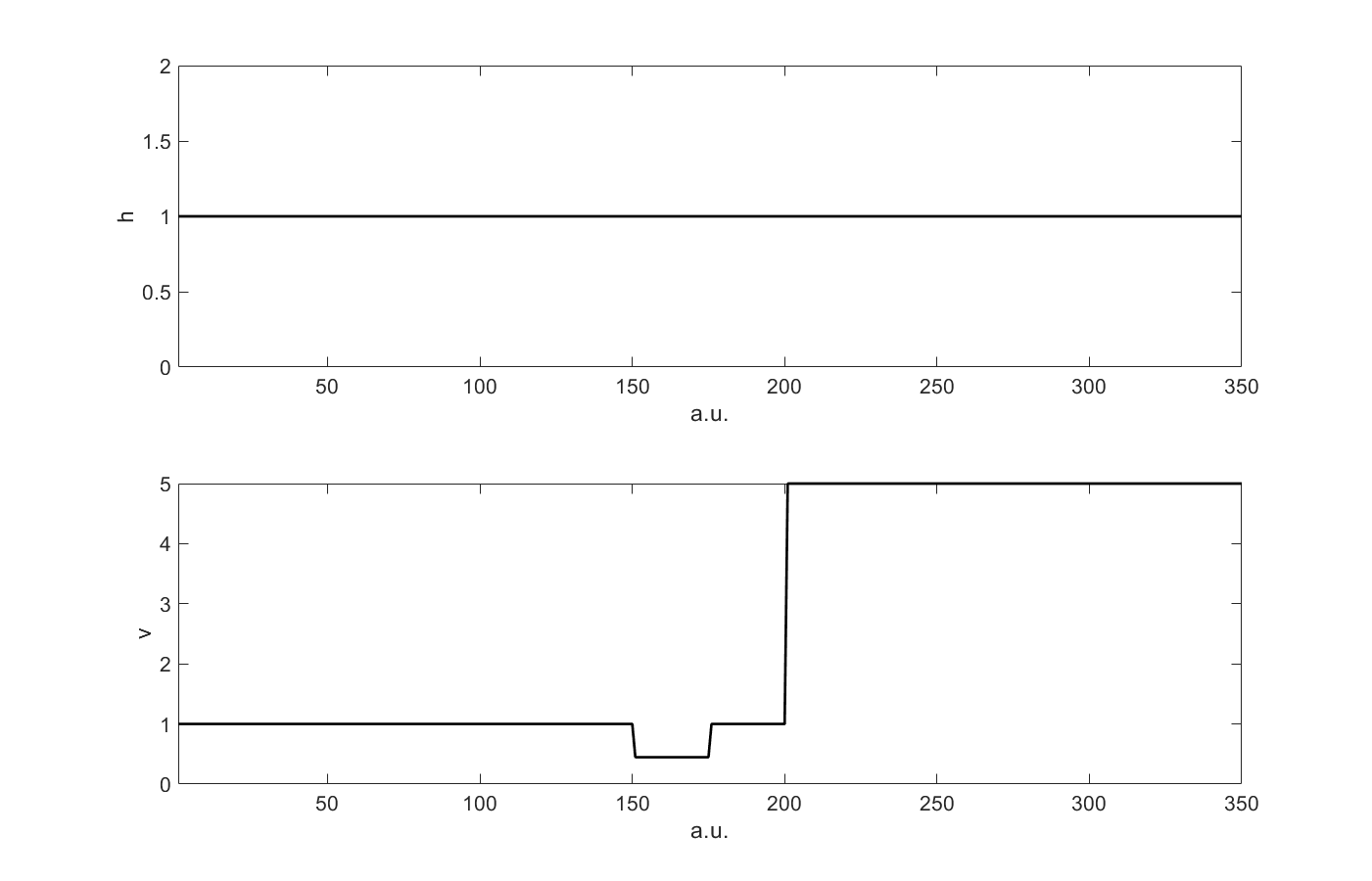}
\includegraphics[width=0.47\textwidth]{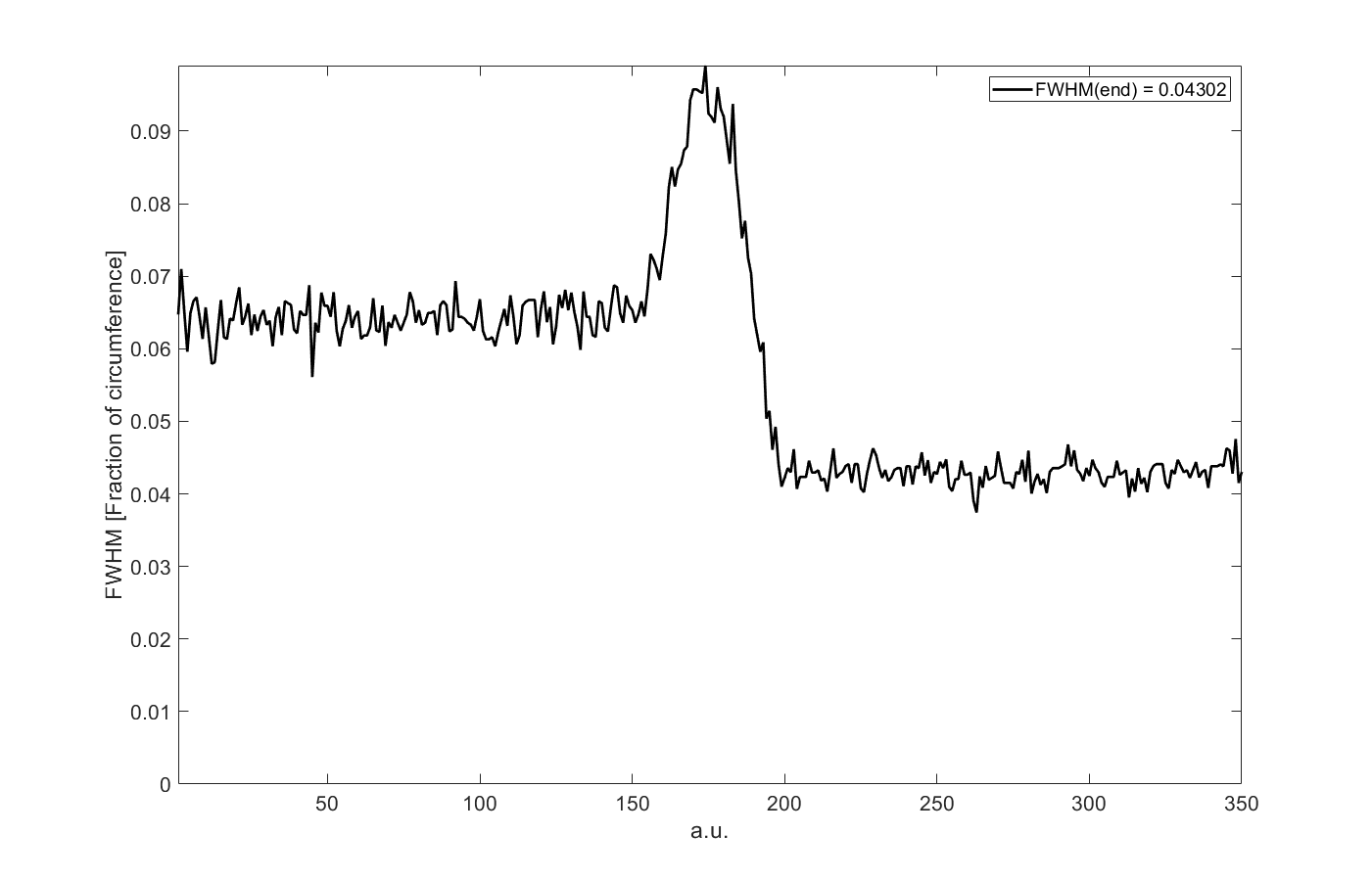}
\end{center}
\caption{\label{fig:transfer4}The evolution of the harmonic (top left), the voltage (bottom left)
  and the bunch length (right) for the matched voltage step.}
\end{figure}
In the first simulation we consider a transition from a first-harmonic system when increasing
the voltage by a factor of five, where we follow 10000 particles of a matched initial distribution
that has an rms bunch length of $10^o$.
\par
The left-hand image in Figure~\ref{fig:transfer4}
shows the evolution of the harmonic number in the upper and of the voltage in the lower panel.
We see that the harmonic does not change from its initial value $h=1$, but the voltage changes
during the munch. First it is reduced and subsequently restored to its previous value, which
is visible prior to increasing the voltage to five times its initial value.
\par
The image on the right-hand side shows the full-width at
half maximum (FWHM) of the distribution along the horizontal axis---the bunch length. Note
that we normalize it to the circumference of the ring. We see that the FWHM is initially
constant, where the displayed time corresponds to 1.5 synchrotron-oscillation periods. Then
the FWHM increases as a consequence of the dropped voltage during the munch and then decreases
to the smaller value, required to obtain a matched distribution at the five times increased
voltage. Then the FWHM does not significantly change any more, indicating that the distribution
is indeed a matched one. The reduction of the FWHM to about 70\,\% of its initial value is
consistent with the value $(v_1/v_2)^{1/4}=(1/5)^{1/4}\approx 0.67$ from Equation~\ref{eq:demag}.
\subsection{Sequence of harmonics}
\begin{figure}[p]
\begin{center}
\includegraphics[width=0.32\textwidth]{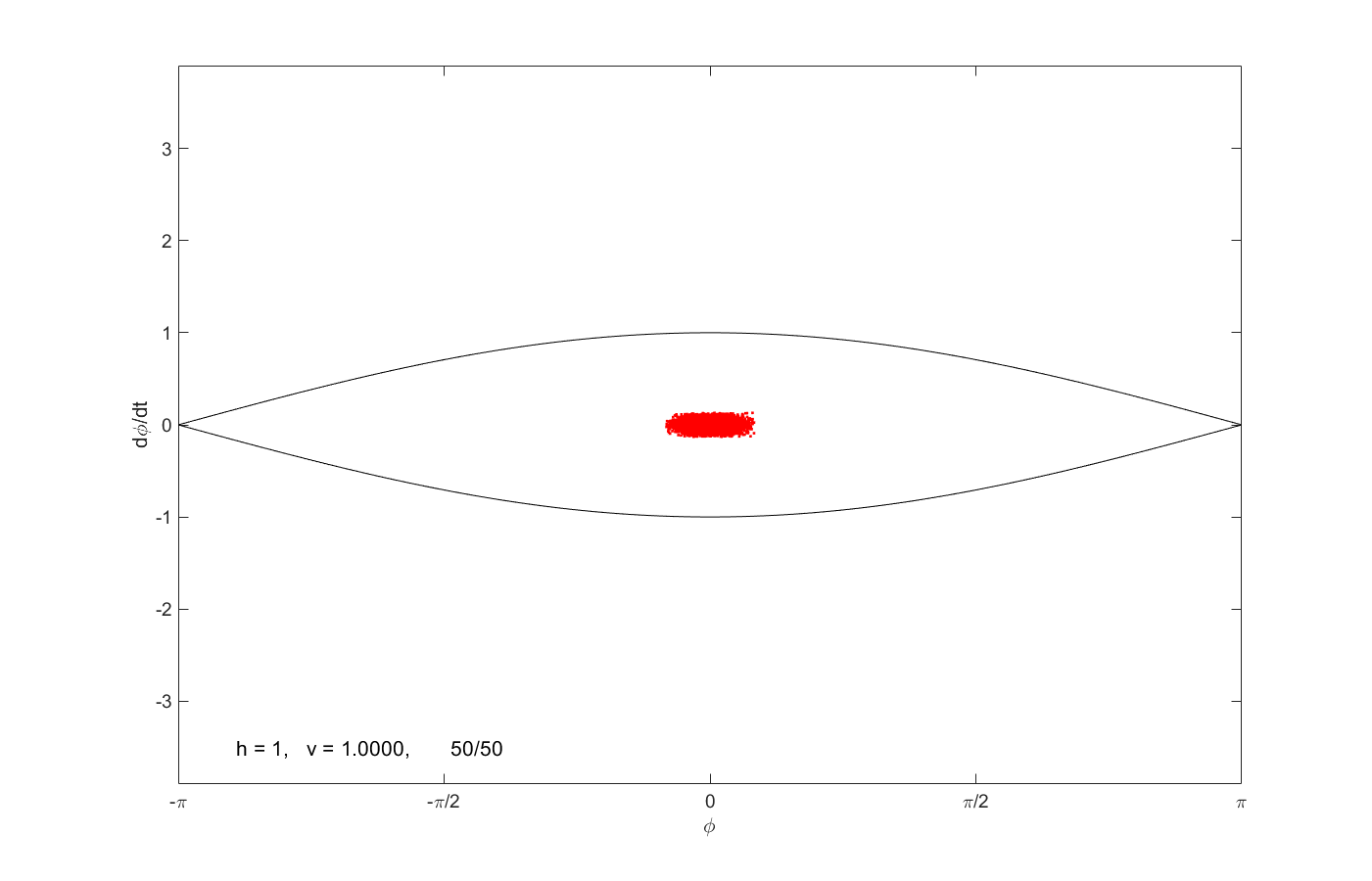}
\includegraphics[width=0.32\textwidth]{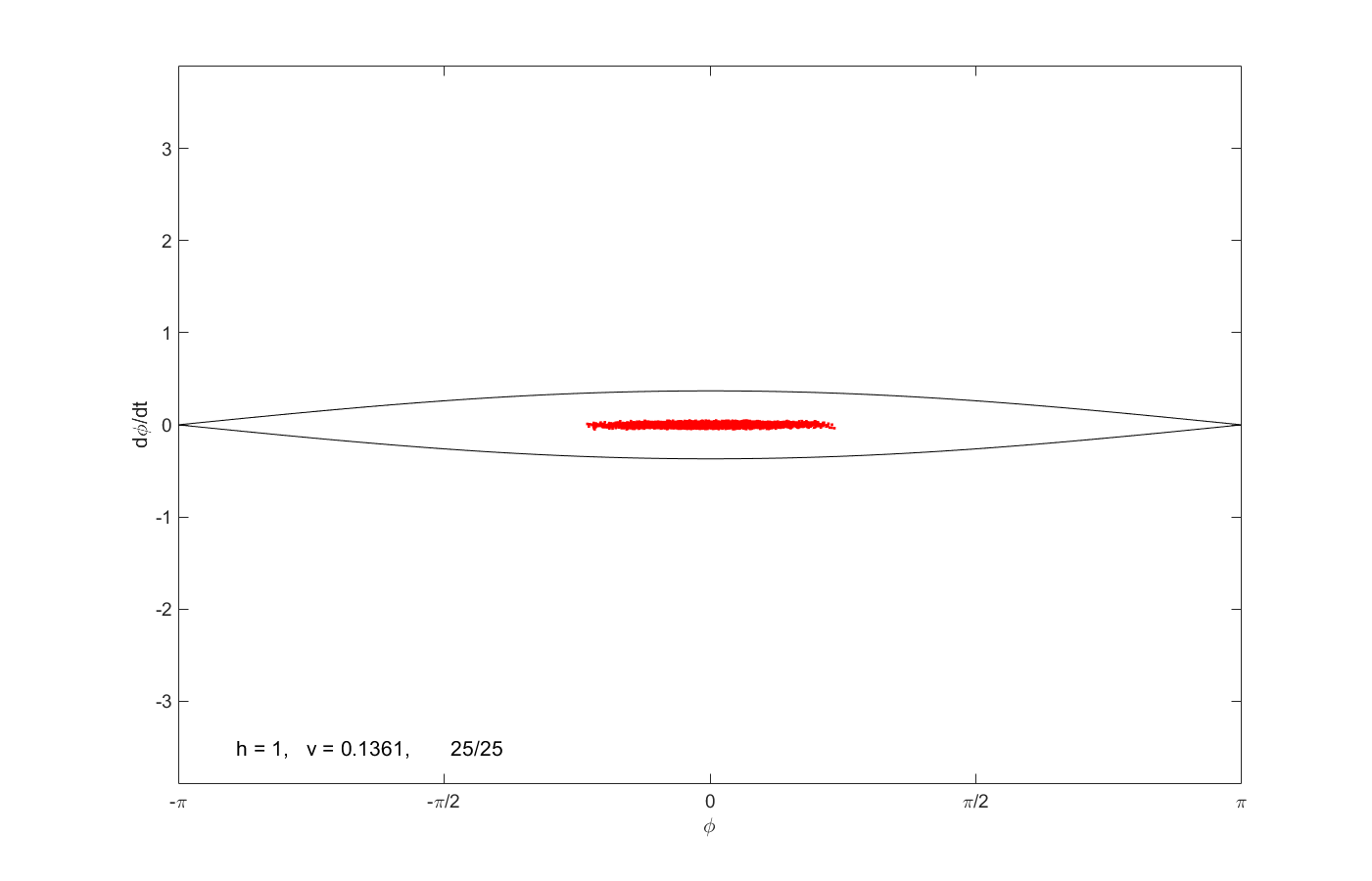}
\includegraphics[width=0.32\textwidth]{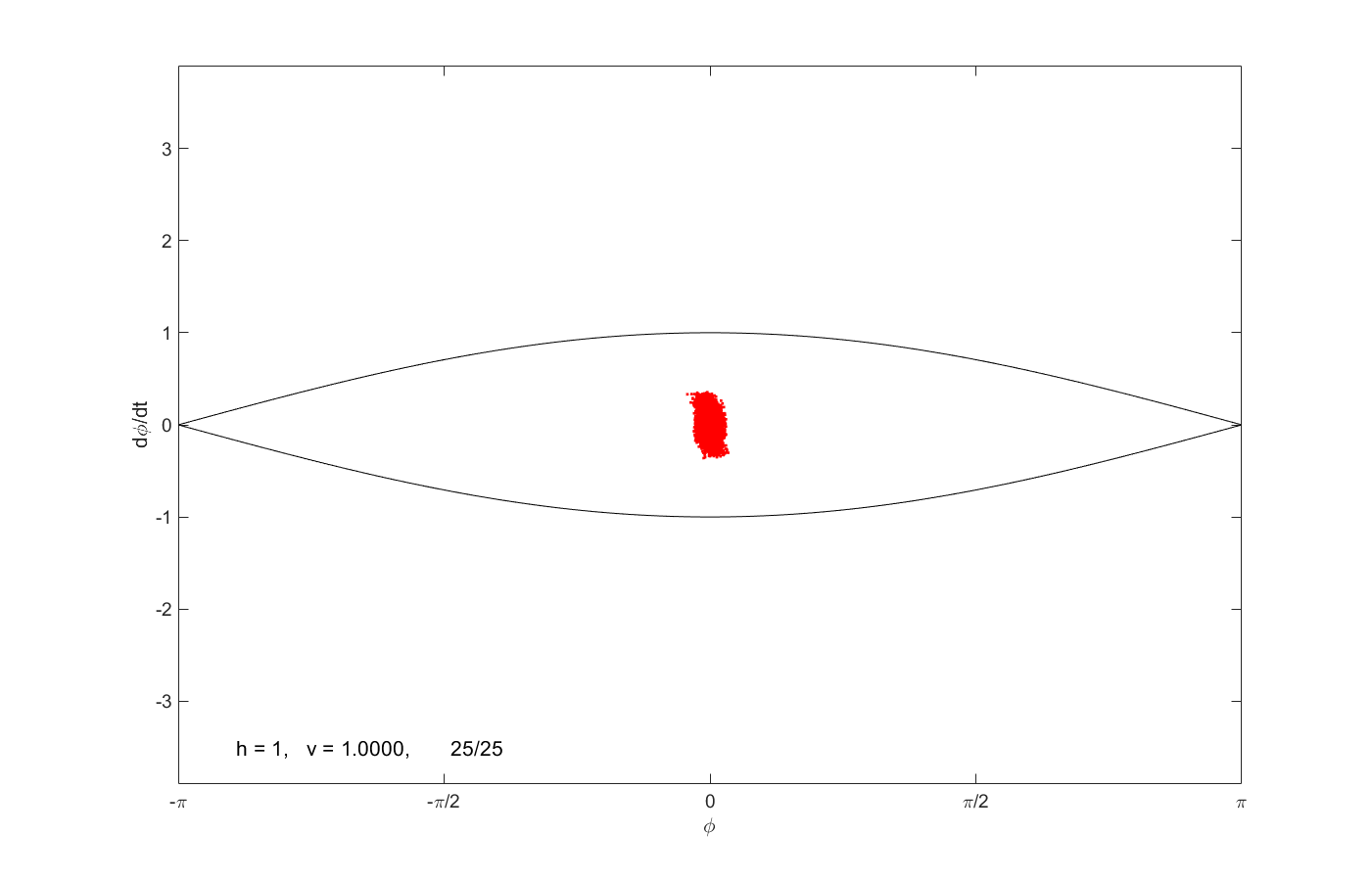}
\includegraphics[width=0.32\textwidth]{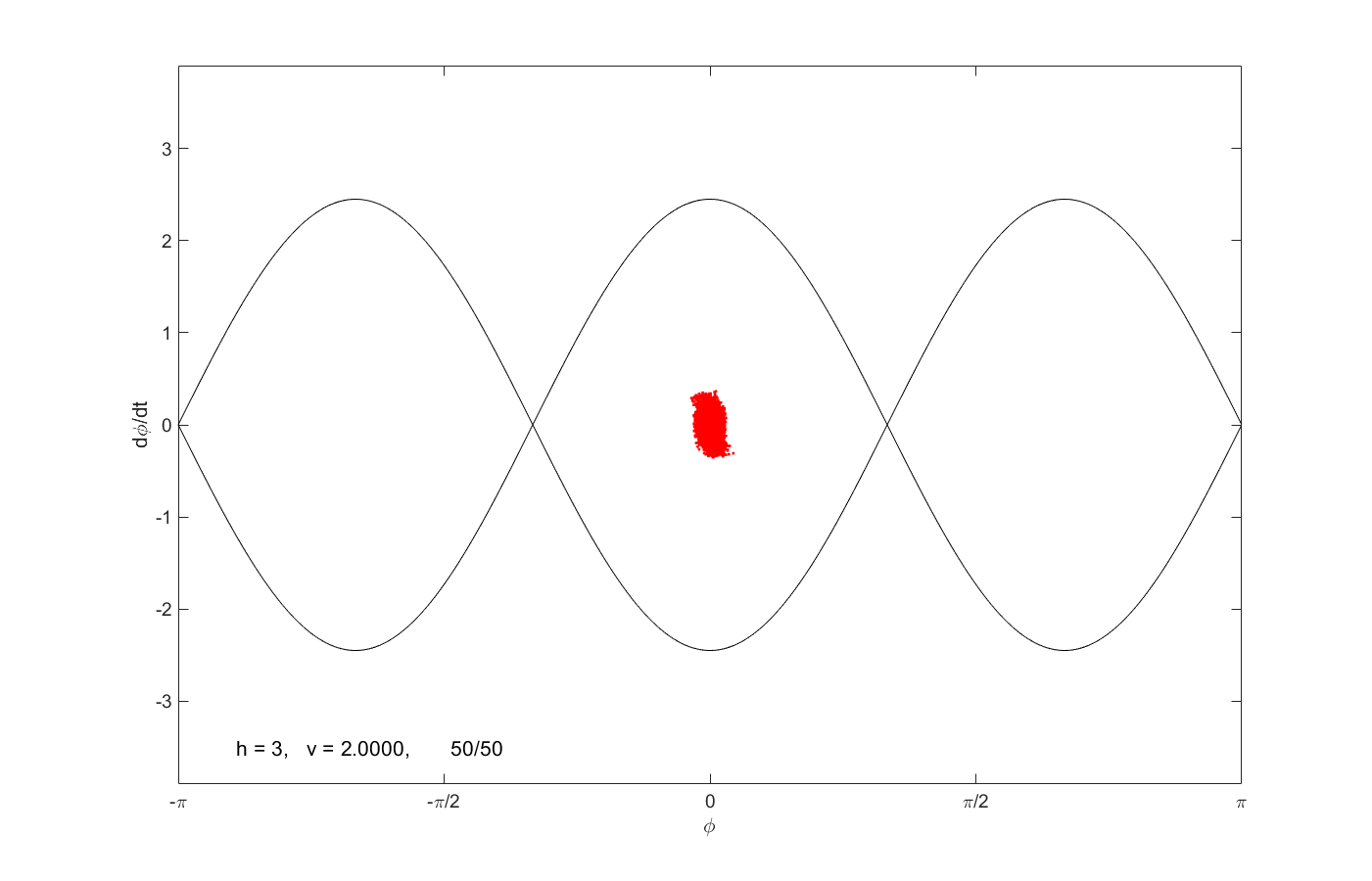}
\includegraphics[width=0.32\textwidth]{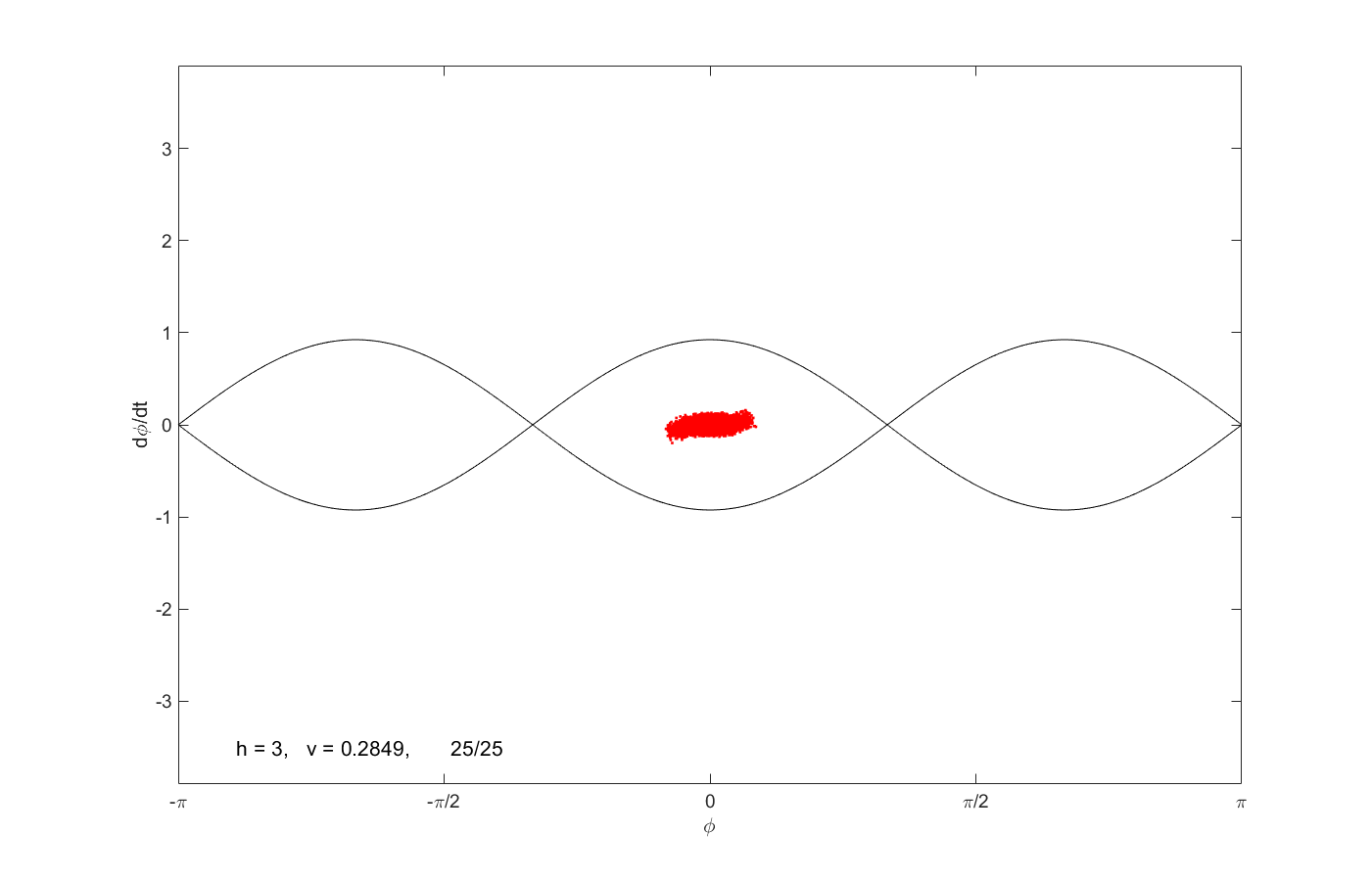}
\includegraphics[width=0.32\textwidth]{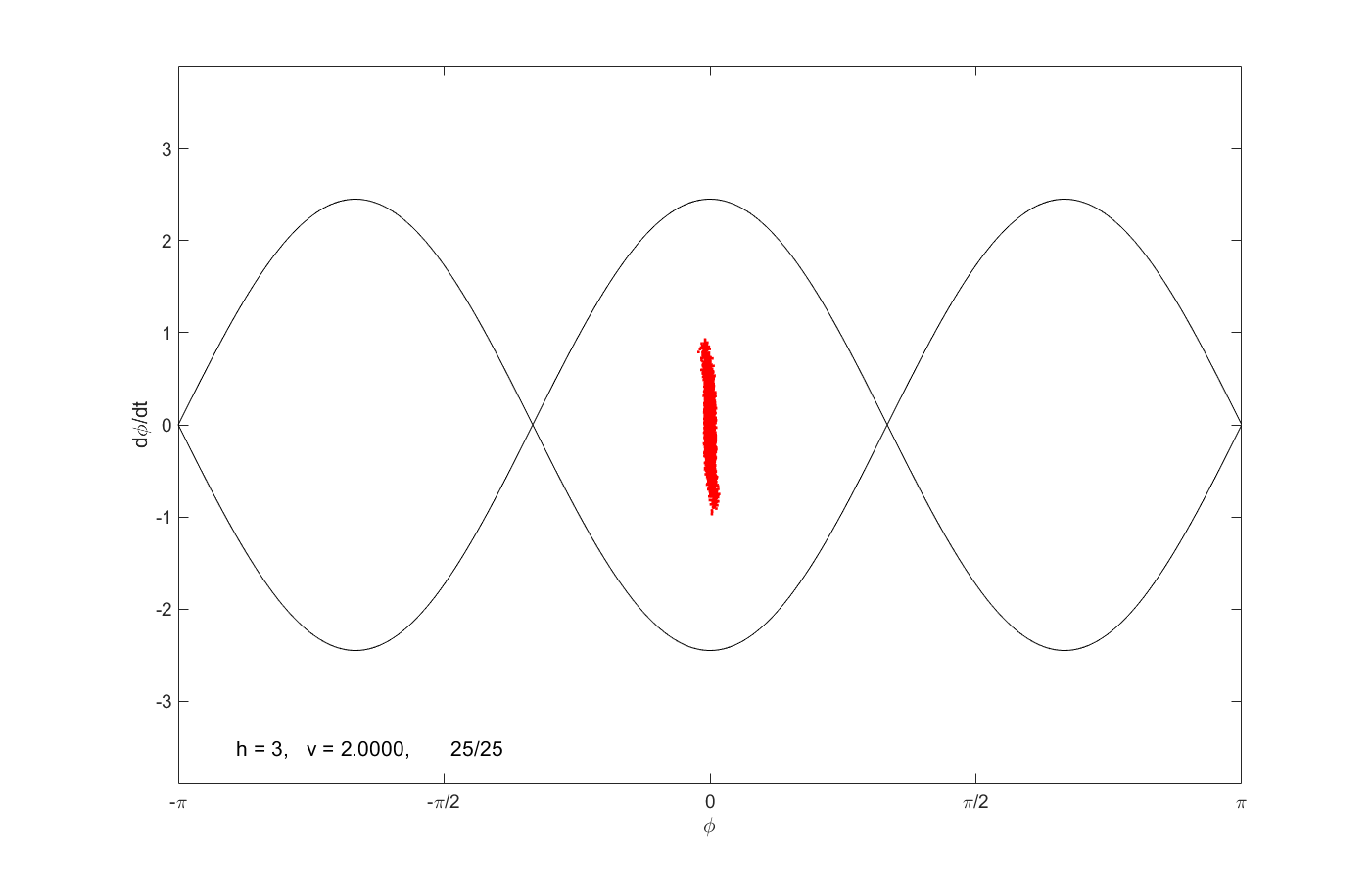}
\includegraphics[width=0.32\textwidth]{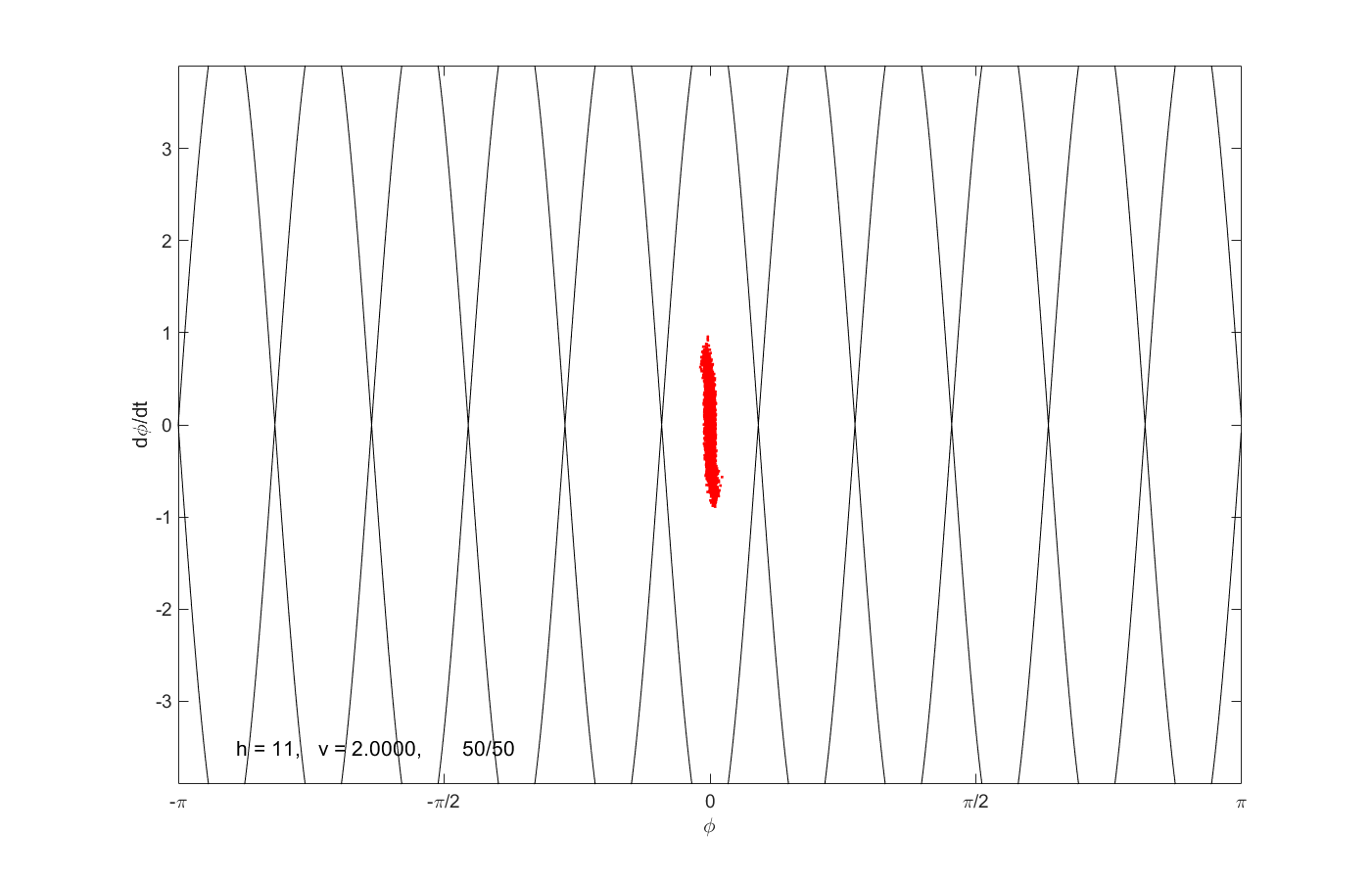}
\includegraphics[width=0.32\textwidth]{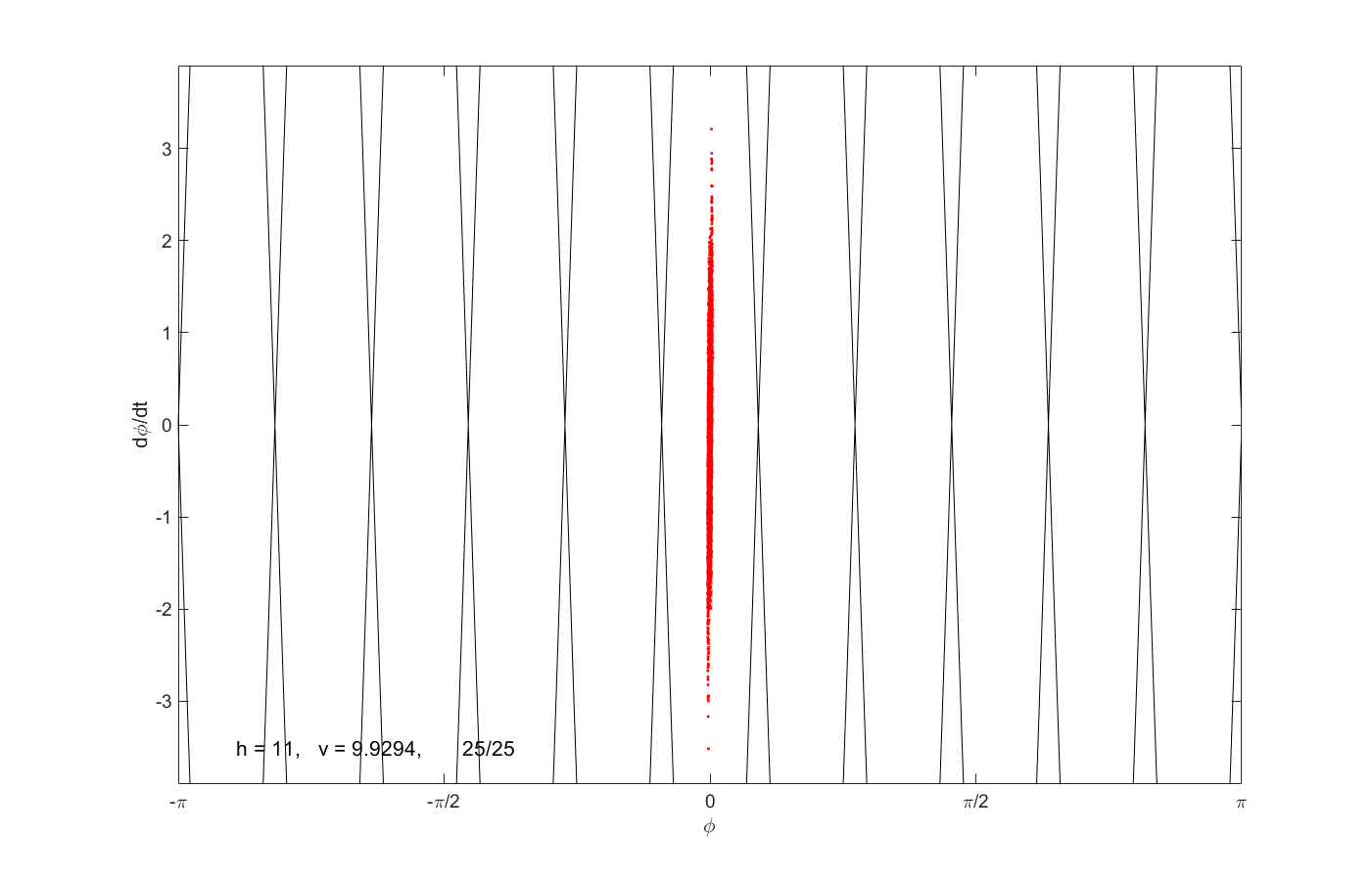}
\includegraphics[width=0.32\textwidth]{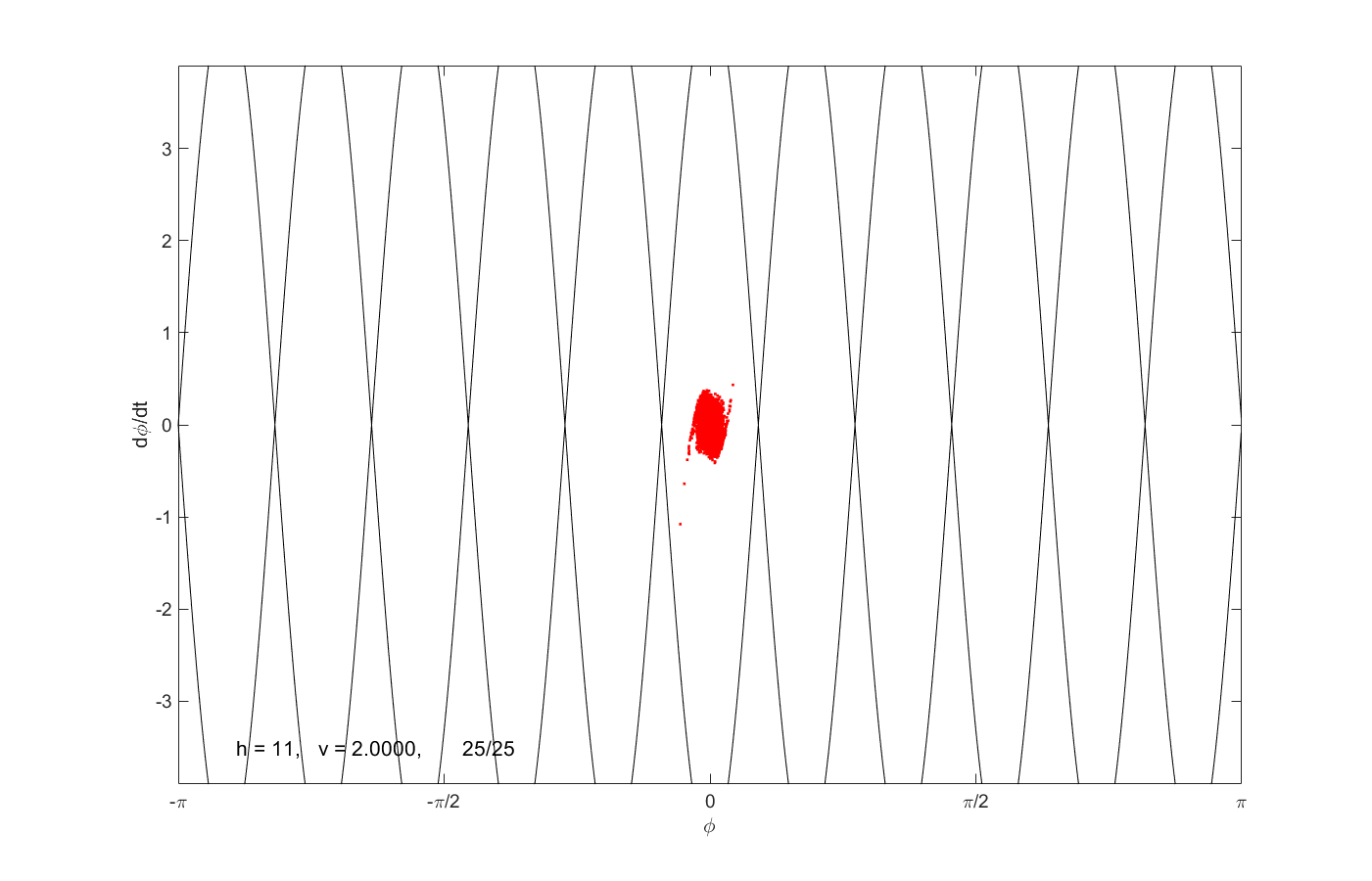}
\includegraphics[width=0.32\textwidth]{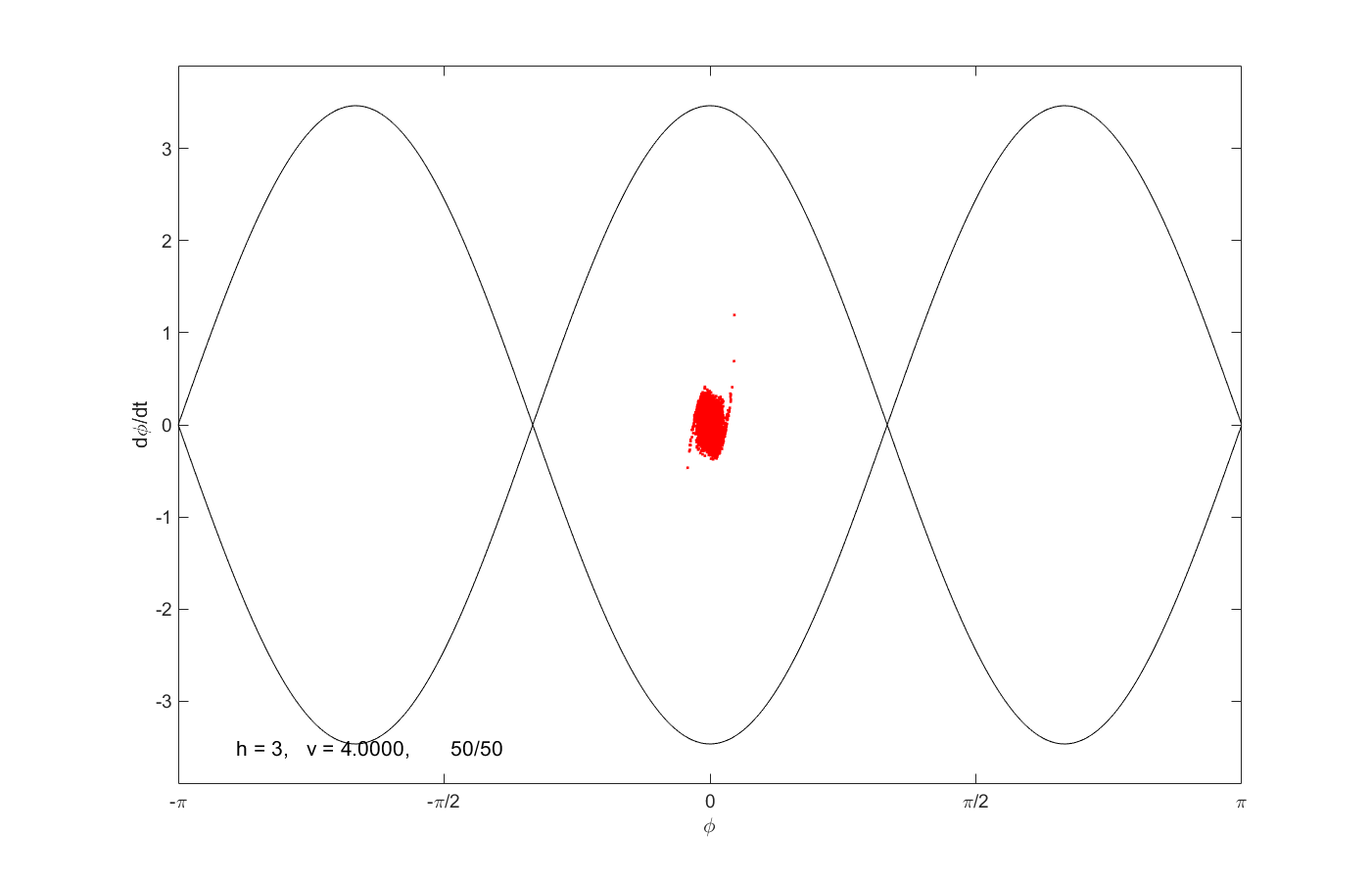}
\includegraphics[width=0.32\textwidth]{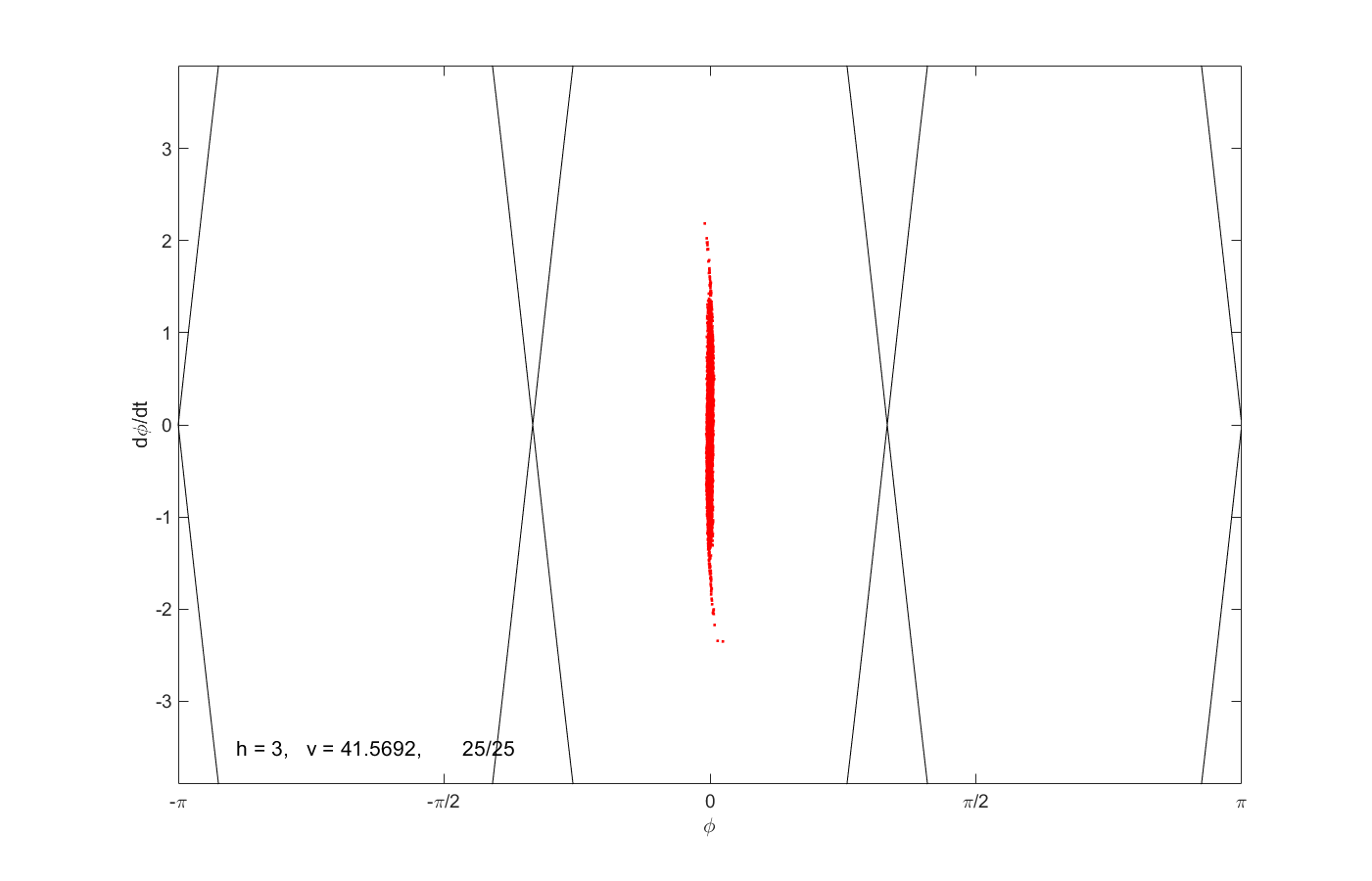}
\includegraphics[width=0.32\textwidth]{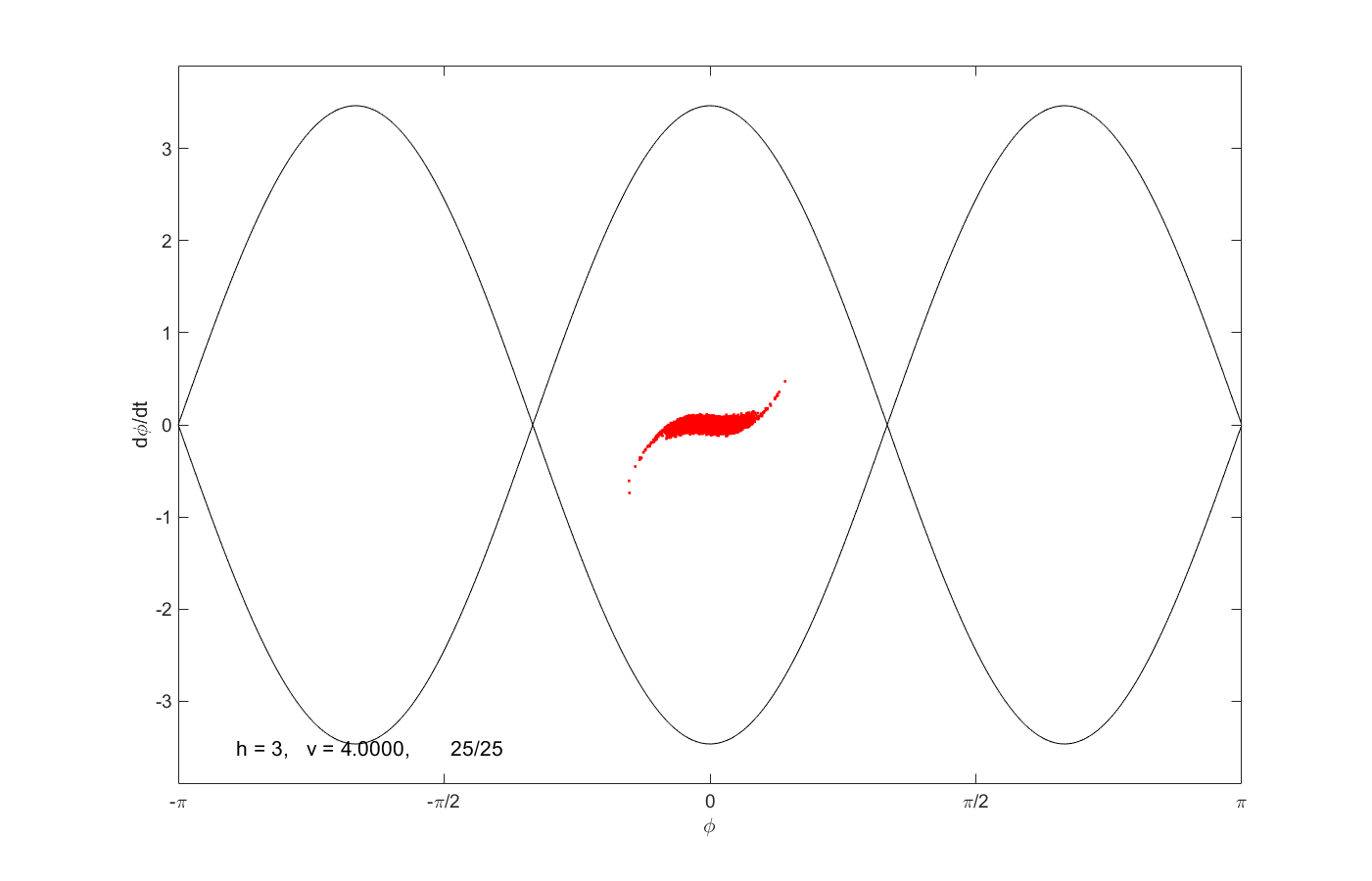}
\includegraphics[width=0.32\textwidth]{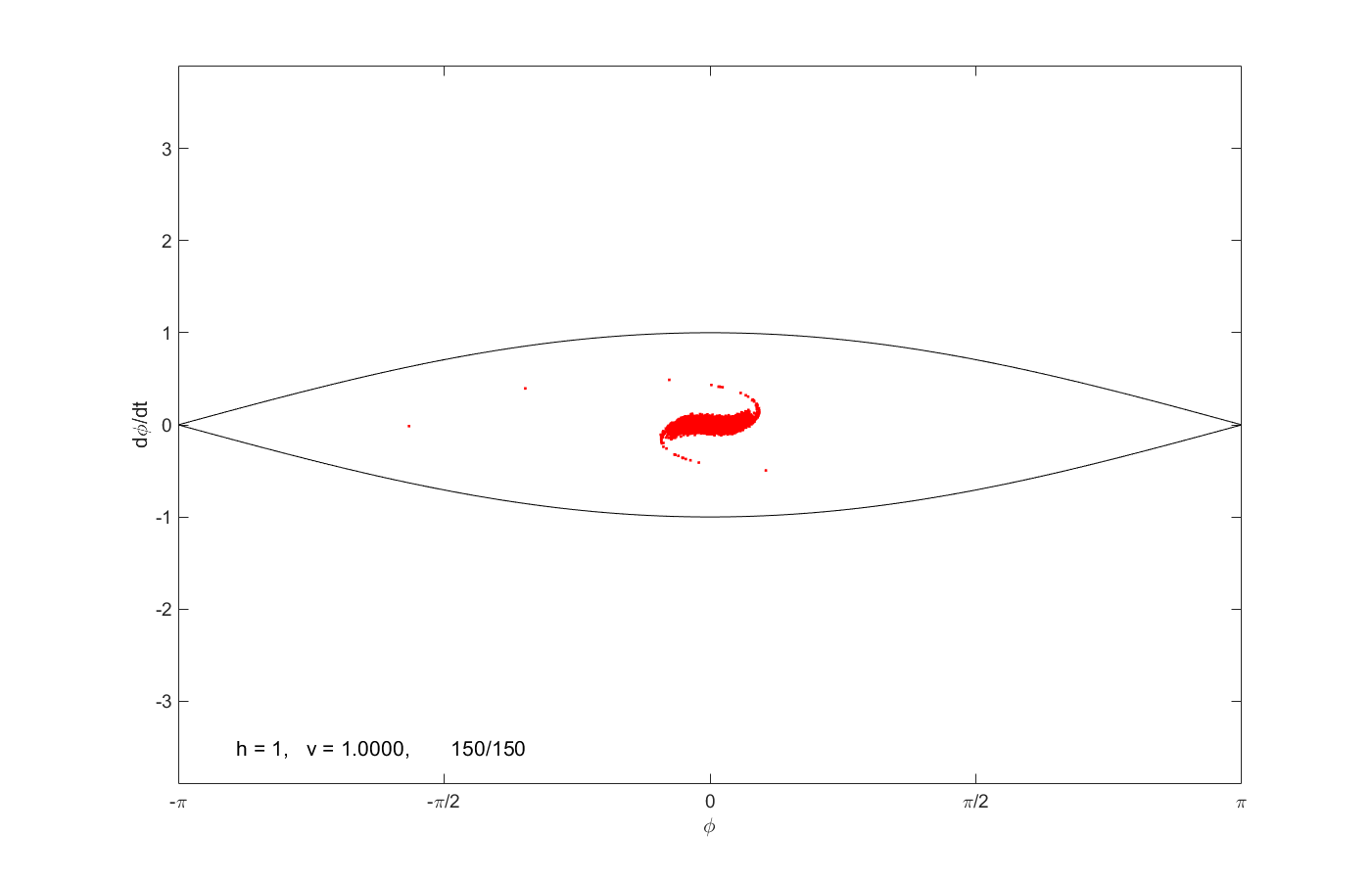}
\includegraphics[width=0.32\textwidth]{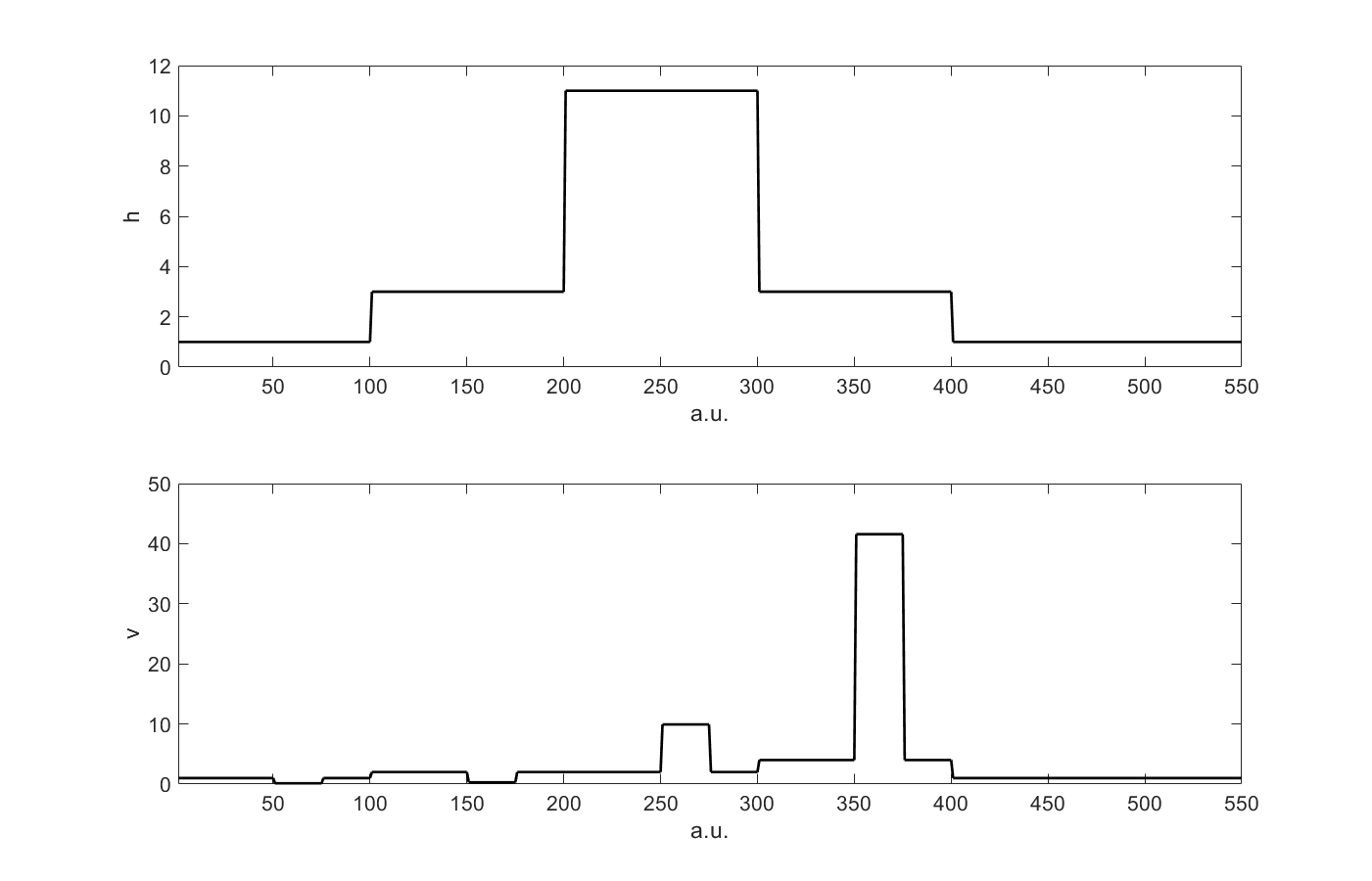}
\includegraphics[width=0.32\textwidth]{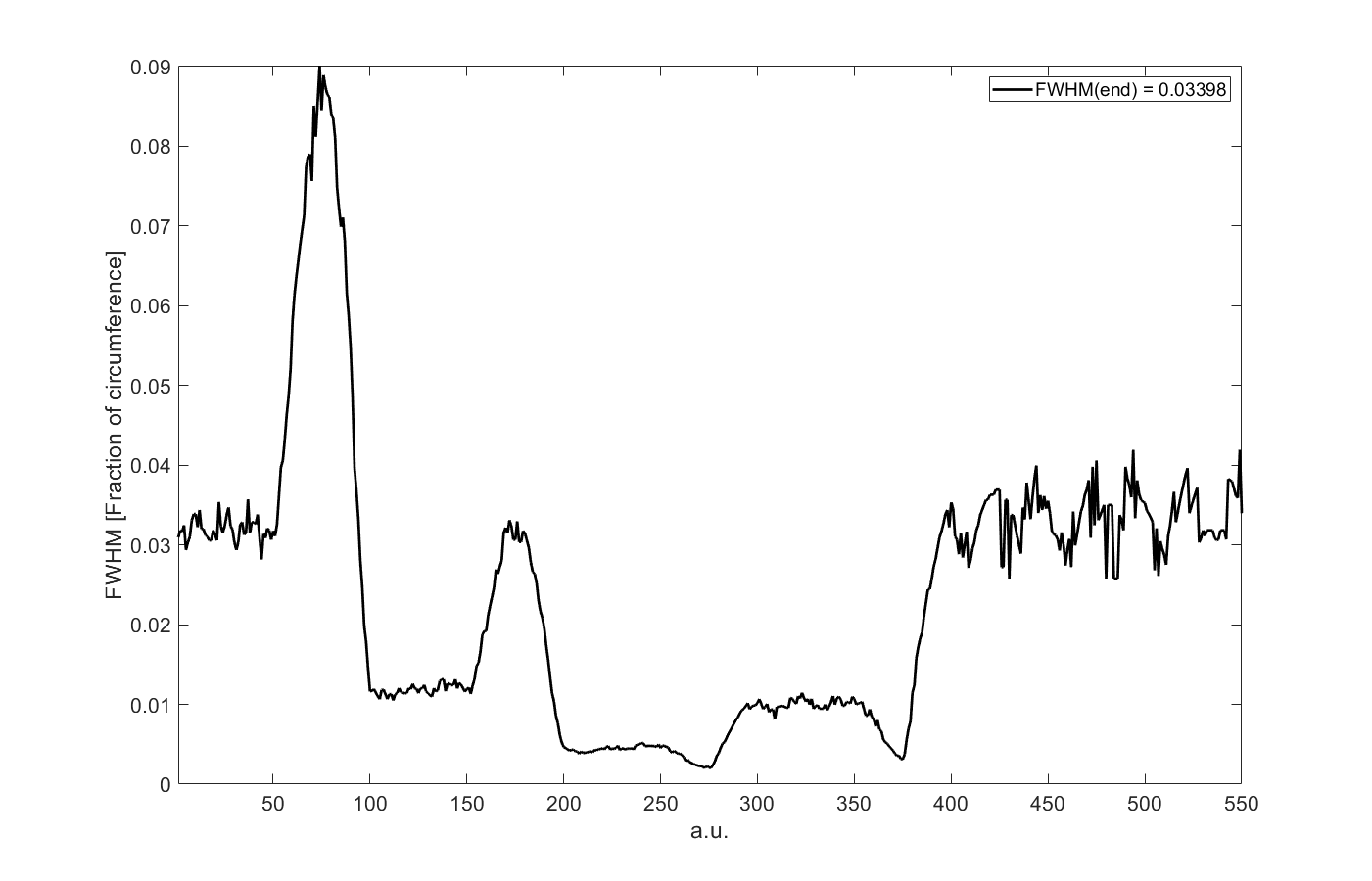}
\end{center}
\caption{\label{fig:transfer1}The phase-space distributions for the $h=1,3,11,3,1$
  sequence of harmonics. In the top row shows the munch to prepare the distribution for
  the third harmonic. The second row shows the munch for the eleventh harmonic, followed
  by the munch back to the third and first harmonic in the third and fourth row. The
  middle plot in the bottom row shows the evolution of harmonic (upper) and voltage
  (lower) and the plot on the right shows the evolution of the bunch length.}
\end{figure}
In the following simulation we transfer the distribution from the first to the third and on to
the eleventh harmonic and then transfer back through similar intermediate stages to the
initial configuration.
\par
We start with a matched distribution of 10000 particles with an rms bunch length of
$10^o$ of the first-harmonic RF system, which is shown on the left plot in the first row in
Figure~\ref{fig:transfer1}. Note that the horizontal axis shows the phase $\phi$, at the
first harmonic such that the shown axis corresponds to the circumference of the ring. The
vertical axis shows $d\phi/dt$ and the black line shows the separatrix. Note that the maximum
height of the separatrix corresponds to the momentum acceptance of the RF system.
\par
Then we create a number of transitions, where we transfer from the first harmonic to 
the third and then to the eleventh harmonic, before returning back via the third to the 
first harmonic. In detail the steps are given in Table~\ref{tab:second}, where the first column 
refers to the position in Figure~\ref{fig:transfer1}, the second and third give the harmonic
$h$ and voltage $v$, whereas the last column contains a description of the state.
\begin{table}[tb]
\begin{center}
\begin{tabular}{c|ccl}
Row, Column & Harmonic $h$  & Voltage $v$  & Comment\\
\hline
1,1 &  1 & 1     & Initial distribution\\
1,2 &  1 & 0.136 & Munch to $h=2,v=3$  \\
1,3 &  1 & 1     & \\
2,1 &  3 & 2     & Third harmonic \\
2,2 &  3 & 0.285 & Munch to $h=11,v=2$  \\
2,3 &  3 & 2     & \\
3,1 & 11 & 2     & Eleventh harmonic \\
3,2 & 11 & 9.93  & Munch back to $h=3,v=4$  \\
3,3 & 11 & 2     & \\
4,1 &  3 & 4     & Third harmonic \\
4,2 &  3 & 42.57 & Munch back to $h=1,v=1$  \\
4,3 &  3 & 4     & \\
5,1 &  1 & 1     & Back at the first harmonic
\end{tabular}
\end{center}
\caption{\label{tab:second}The steps in the second simulation though the sequence of harmonics.}
\end{table}
In Figure~\ref{fig:transfer1} we show the bunch distribution at the end of each state described
in Table~\ref{tab:second}. In the first row the three images show the initial distribution on 
the left, the distribution at the end of the munch and on the right after the quarter-period
rotation with the restored voltage. At this point the bunch is still governed by the first
harmonic, but already 
has the shape of the matched distribution of the third harmonic. At this point we turn off 
the first-harmonic system, turn on the third-harmonic system, and follow the distribution
for half a synchrotron period. It is shown on the first image in the second row, which
is practically the same distribution as the one we handed over to the third. At this point
we lower the voltage of the third harmonic---the second munch---and show the stretched
distribution on the image in the middle of the second row, followed by the distribution
after the quarter oscillation with restored voltage, shown on the right. At this point the 
distribution is matched to the eleventh harmonic and we turn off the third harmonic and 
turn on the eleventh. The distribution after half a synchrotron oscillation is shown on the
left image in the third row.
\par
At this point we decide to reverse the sequence. In the first stage we have to munch back to
the third harmonic and that entails to briefly, albeit significantly, increase the RF voltage 
of the eleventh, which is shown in the middle image in the third row. Subsequently restoring
the voltage to the original value for a quarter period turns the distribution to a matched
distribution suitable for the third harmonic (at a different voltage than before). After 
turning off the eleventh-harmonic system and turning on the third we show the distribution 
after half a synchrotron period in the first image in the fourth row. Next we increase the
voltage, shown in the middle image, and after  a quarter oscillation with restored voltage
the distribution is suitable for the first harmonic, shown on the right. Turning off the
third harmonic and turning on the first, we follow the distribution for 1.5 synchrotron 
periods after which it is shown on the left image in the bottom row. It looks very similar
to the original distribution on the top-left image, only some spiral arms have developed
because in some intermediate state, particles moved outside the linear regime of the 
pendulum equation. The upper panel on the image in the middle of the bottom row shows the 
evolution of the harmonic~$h$ as the simulation progressed and the lower panel shows
the voltage~$v$. Here we note that in particular the voltage during the step-down sequence
reaches prohibitively large values, which are likely to be impossible in a real machine.
That the simulation
nevertheless shows that the method works can be seen from the image on the bottom right in
Figure~\ref{fig:transfer1}. It shows the FWHM of the distribution---the bunch length.
We find that the final FWHM is only slightly larger and more noisy than the initial
value. We attribute the noisiness to the spiral arms of the final distributions. Note also 
that the  munches where the bunch is lengthened show up as spikes on the plot and that 
the smallest values of the FWHM are below one percent of the circumference. For a 300\,m 
ring this is in the tens of ns range. 
\par
We can draw a number of conclusions. First, the steps preserve the longitudinal phase-space 
area, the emittance, and they are almost reversible. Second, going from  a small harmonic
$h_1$  to a larger $h_2$ requires a reduced voltage during  the munch, whereas going the other 
way requires an increased voltage. Third, the beam distribution must stay well inside the 
stable phase-space area and preferably in the linear region. This is difficult when covering
large steps in the harmonics $|h_1-h_2|\gg 1$. If particles end up in the non-linear region,
they will develop into the spiral arms, which increase the emittance and cause the sequence 
to become irreversible. This implies that the phase-space area occupied by the beam, its
longitudinal emittance, must be significantly smaller than the area of the stable phase-space
region, also known as the bucket. Conversely, if the initial emittance is comparable to
the size of the bucket, we cannot munch the bunch towards higher harmonics and shorter
bunch lengths.
\par
Nevertheless we explore a gentle path towards higher harmonics in which we always double the
harmonic in the following section.
\subsection{Harmonic-doubling cascade}
\begin{figure}[p]
\begin{center}
\includegraphics[width=0.3\textwidth]{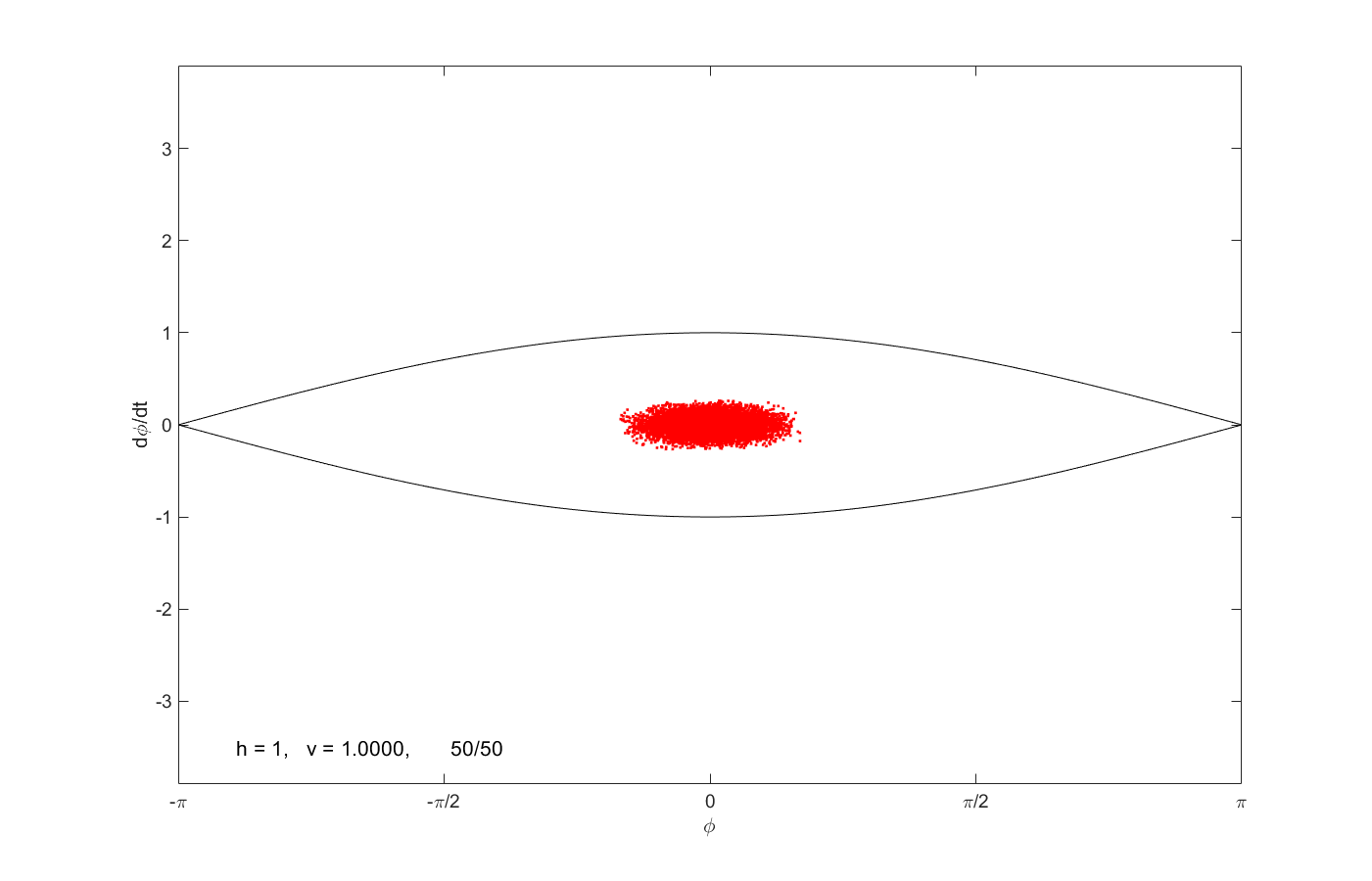}
\includegraphics[width=0.3\textwidth]{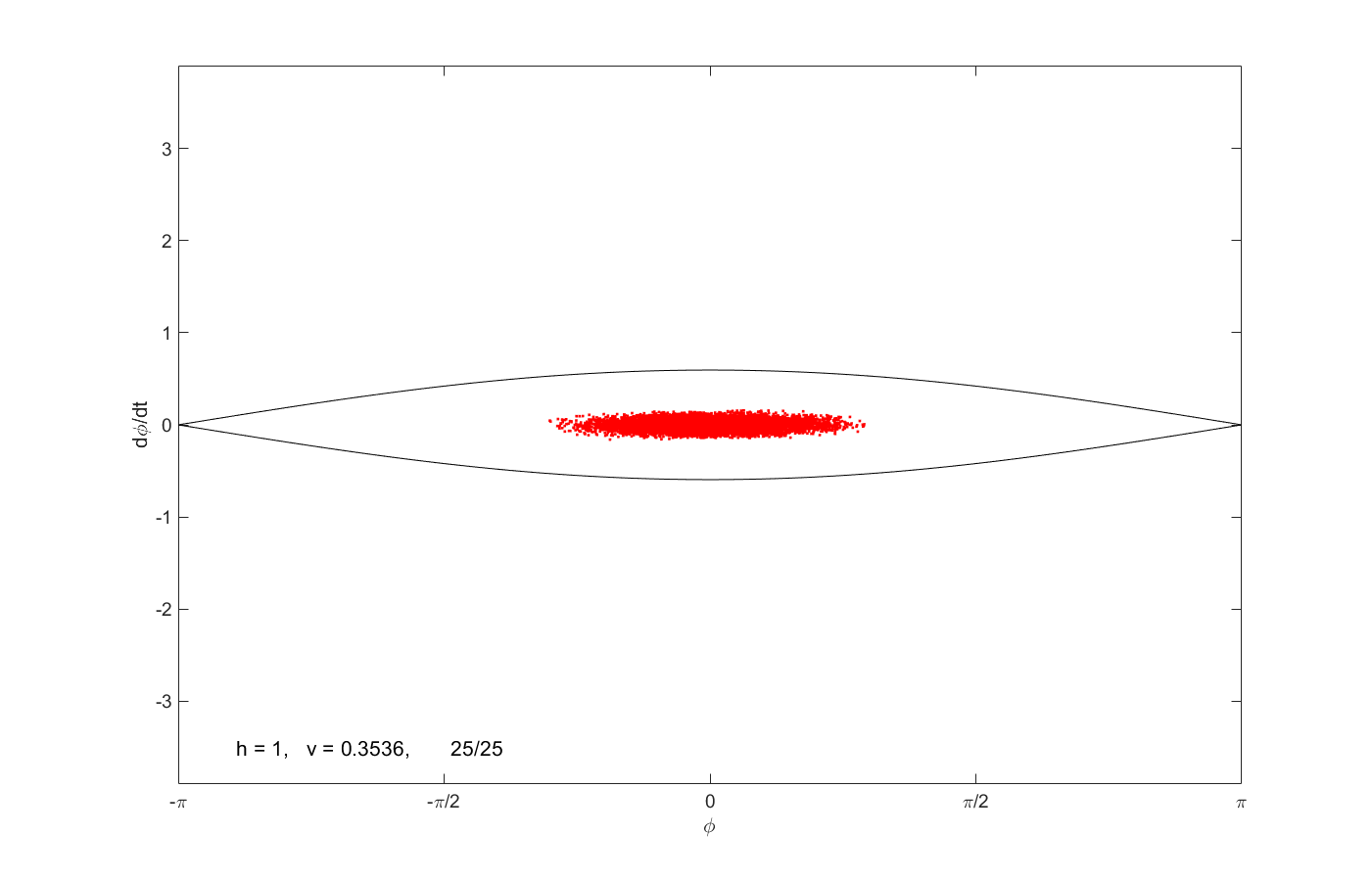}
\includegraphics[width=0.3\textwidth]{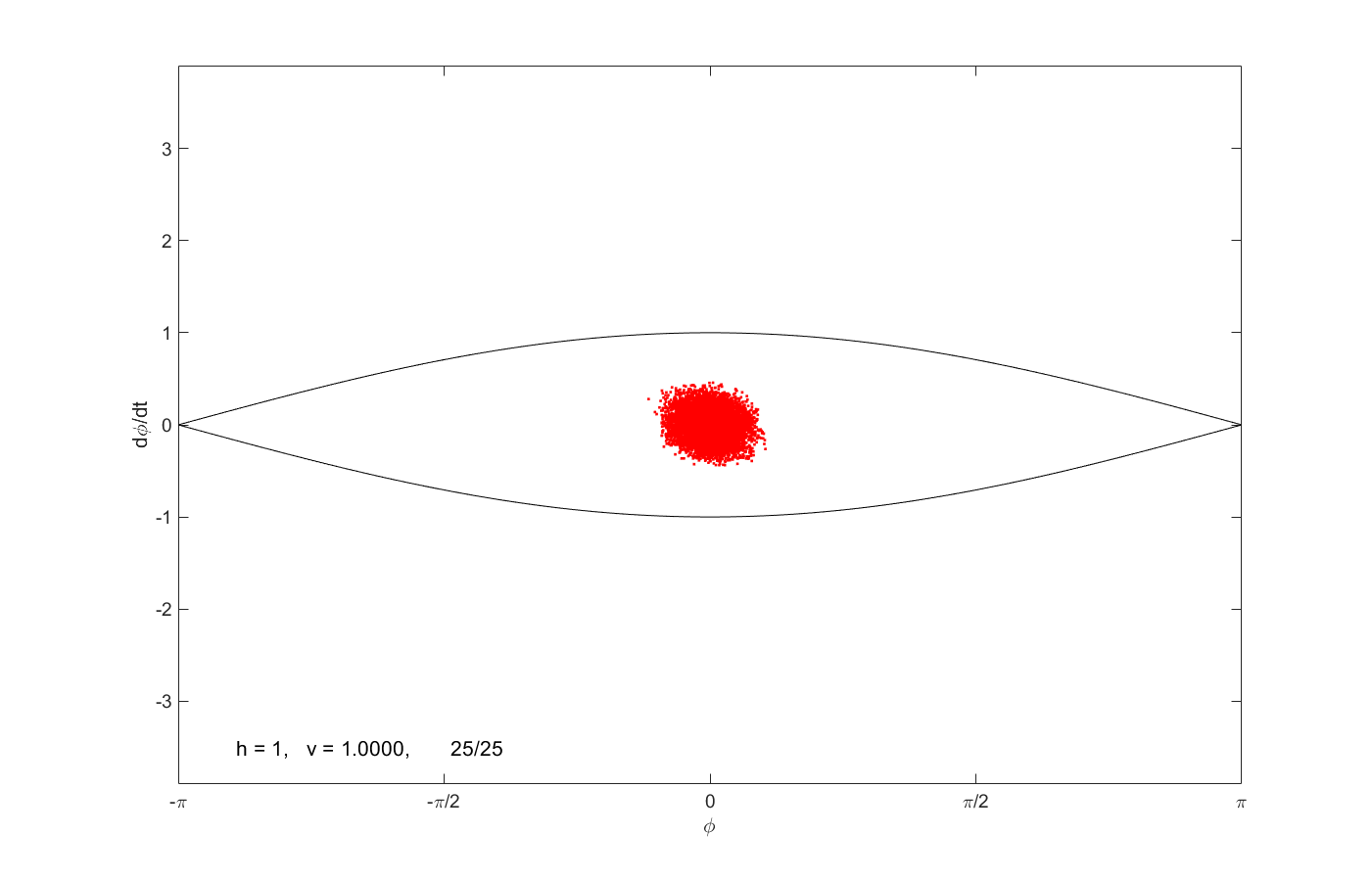}
\includegraphics[width=0.3\textwidth]{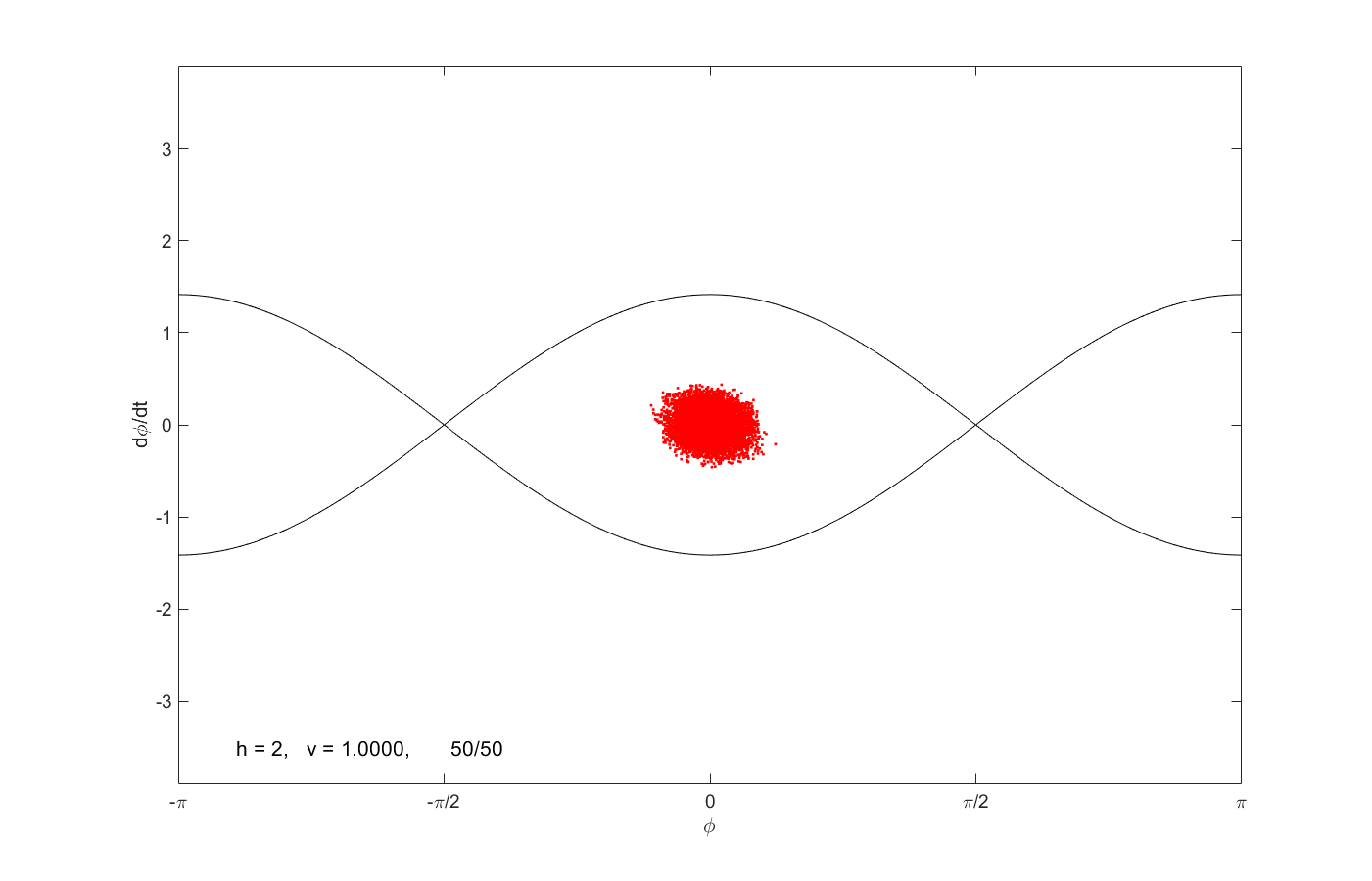}
\includegraphics[width=0.3\textwidth]{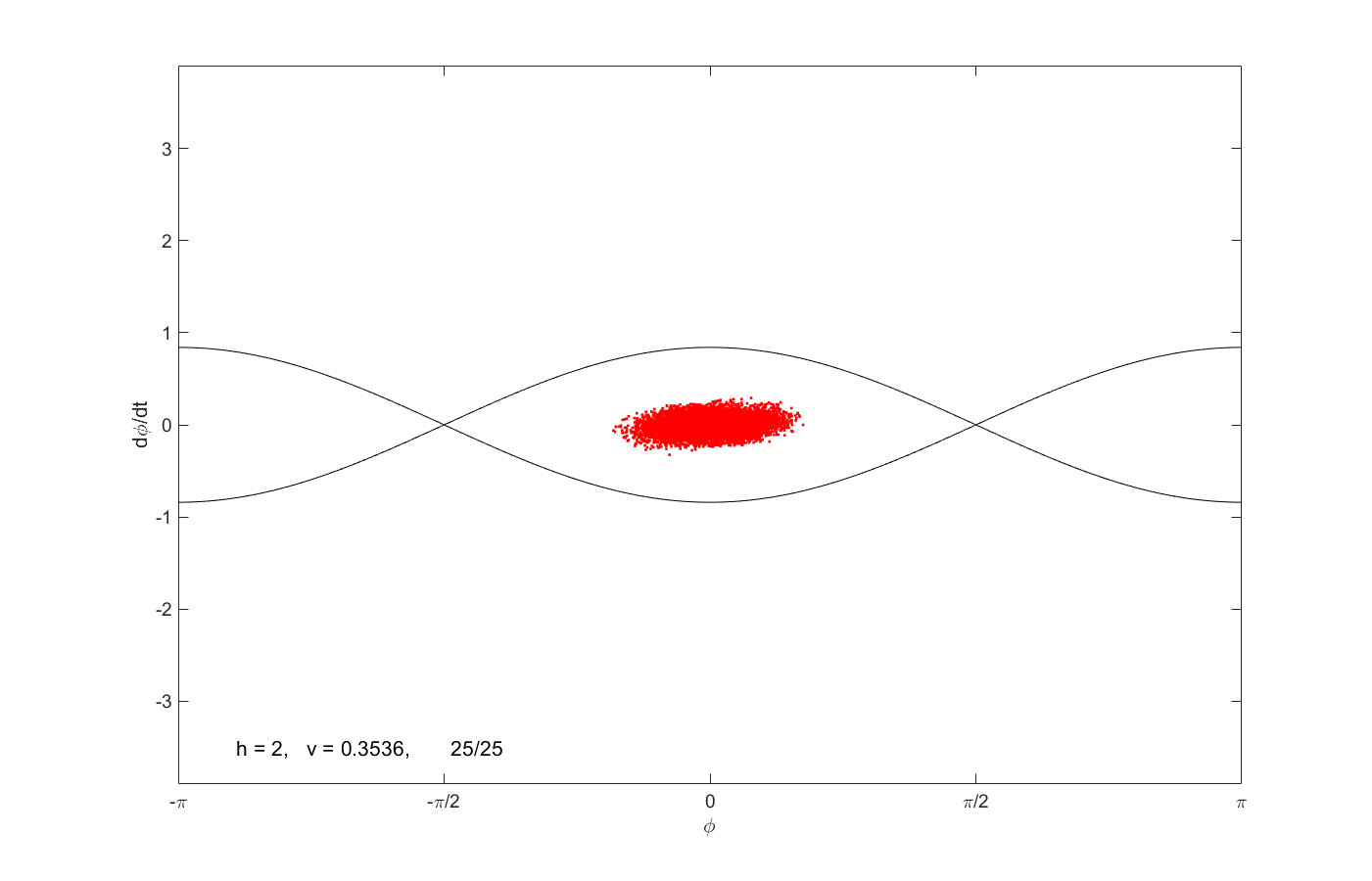}
\includegraphics[width=0.3\textwidth]{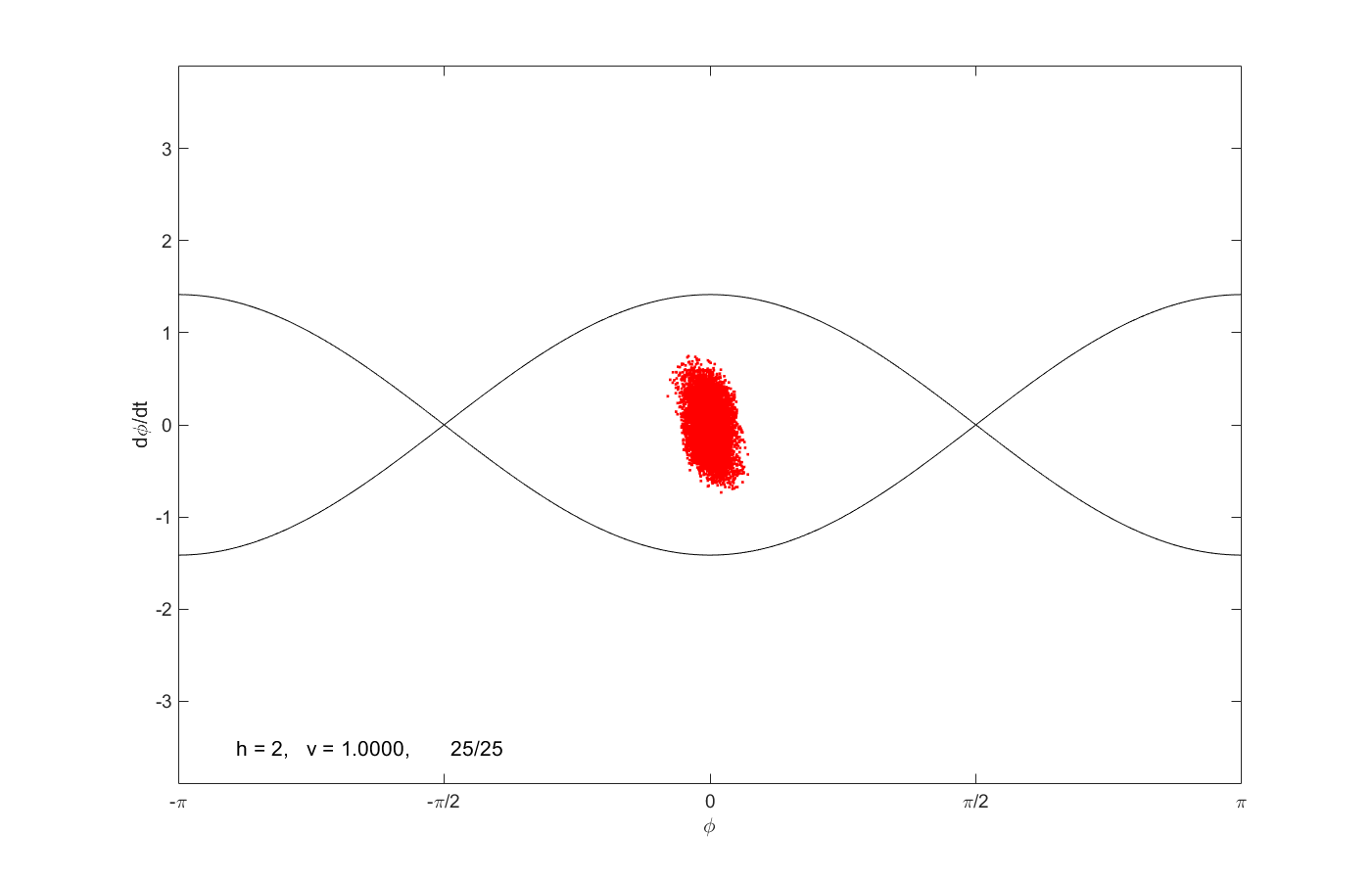}
\includegraphics[width=0.3\textwidth]{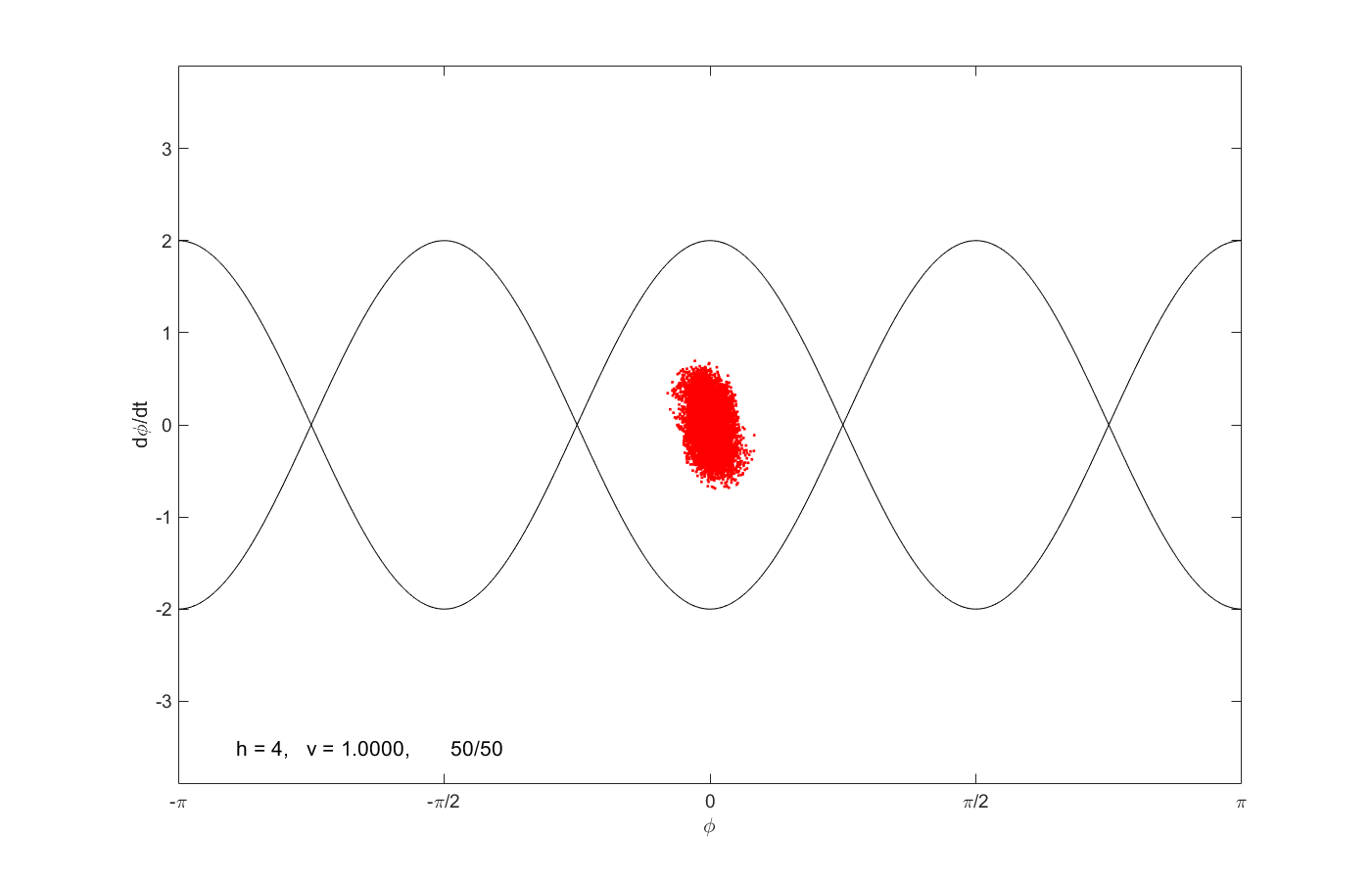}
\includegraphics[width=0.3\textwidth]{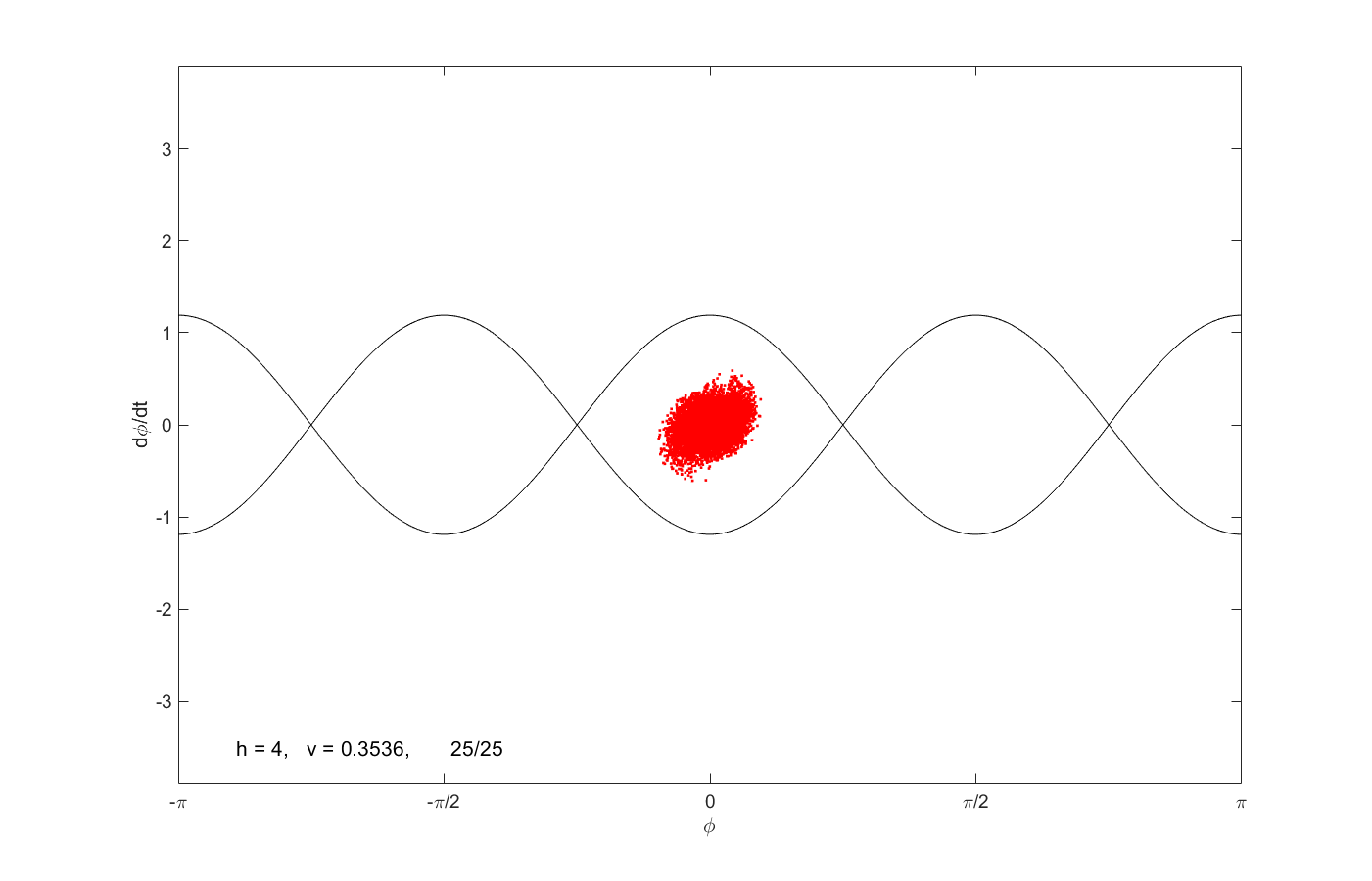}
\includegraphics[width=0.3\textwidth]{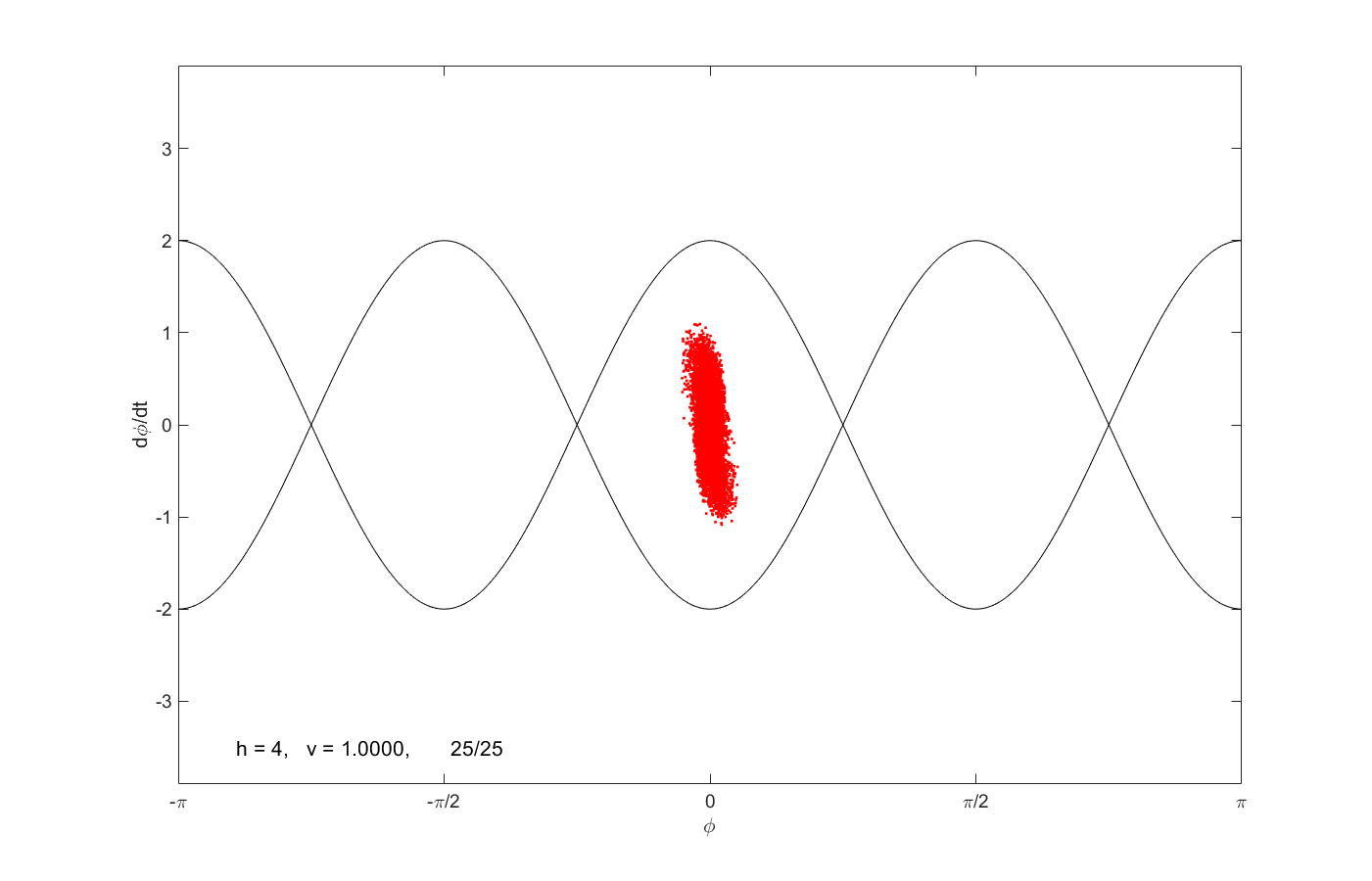}
\includegraphics[width=0.3\textwidth]{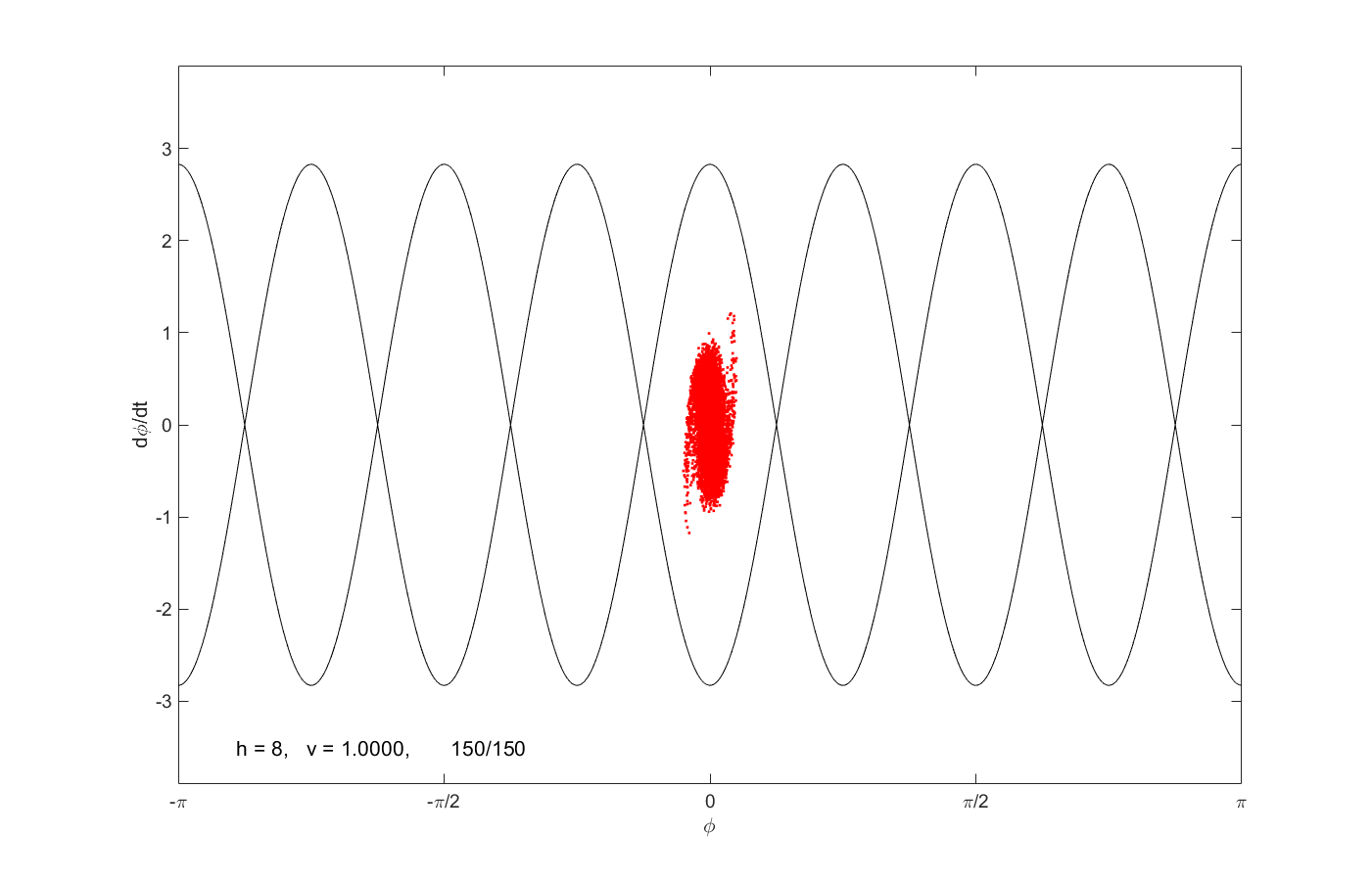}
\includegraphics[width=0.3\textwidth]{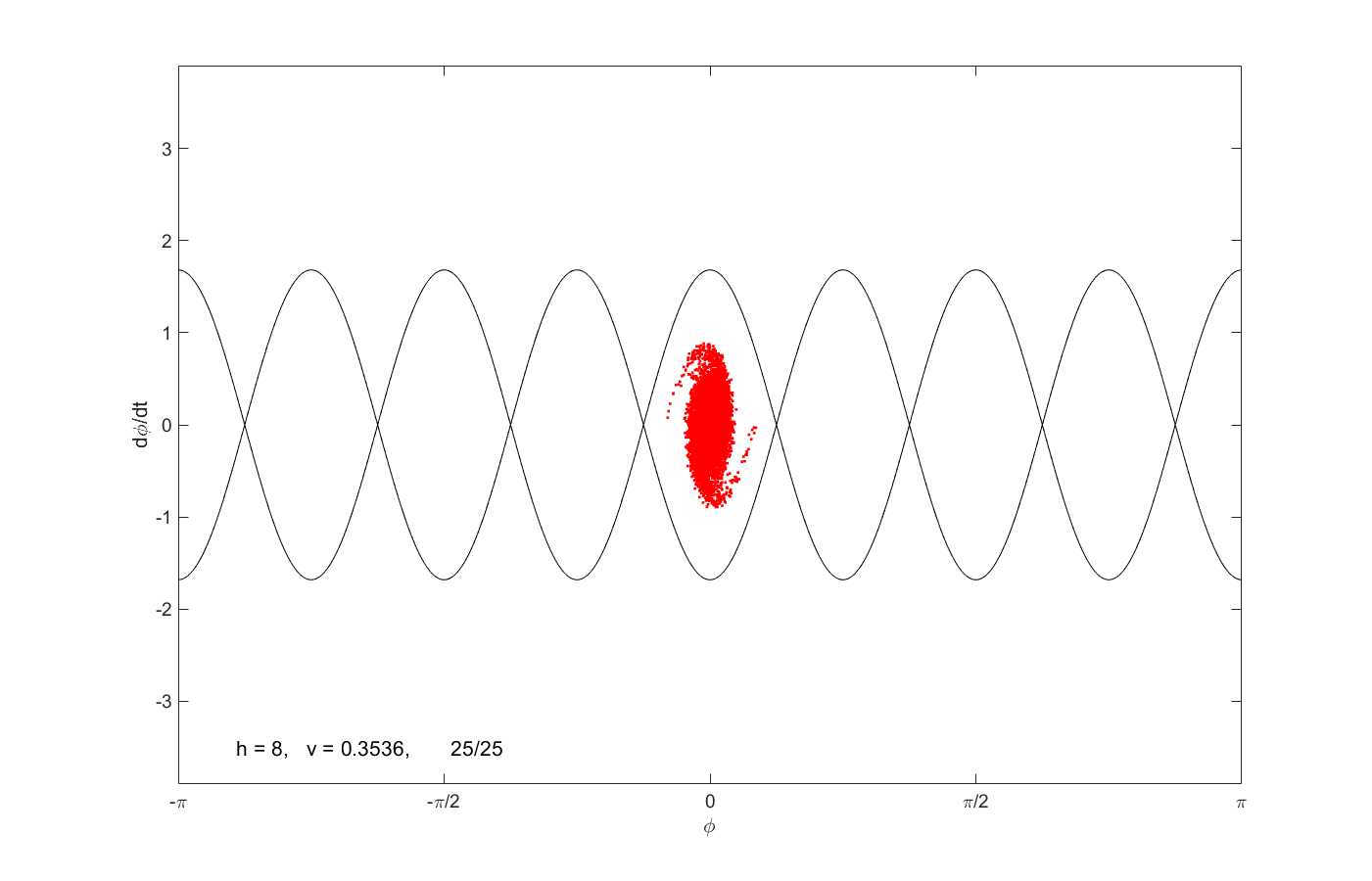}
\includegraphics[width=0.3\textwidth]{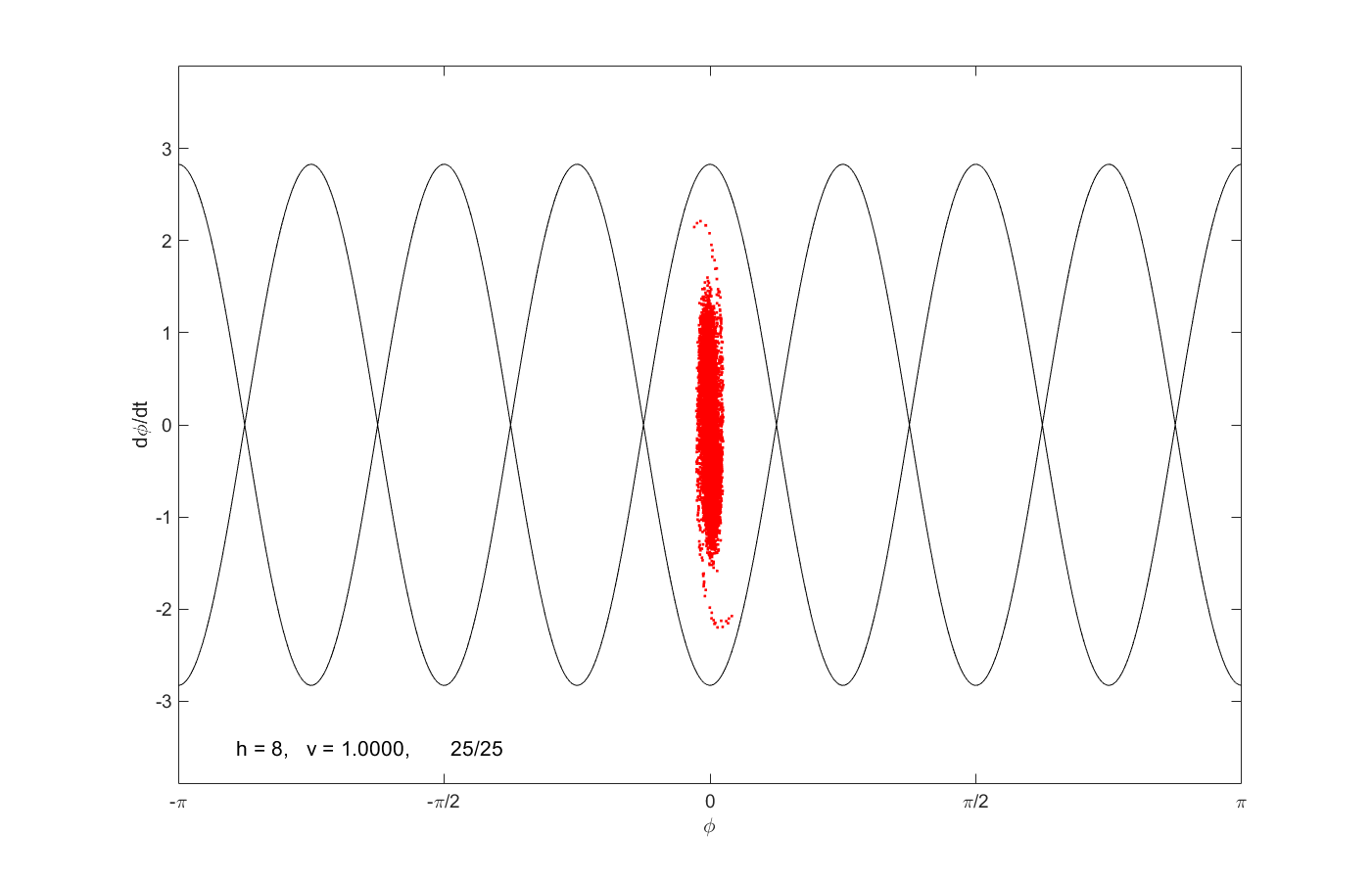}
\includegraphics[width=0.3\textwidth]{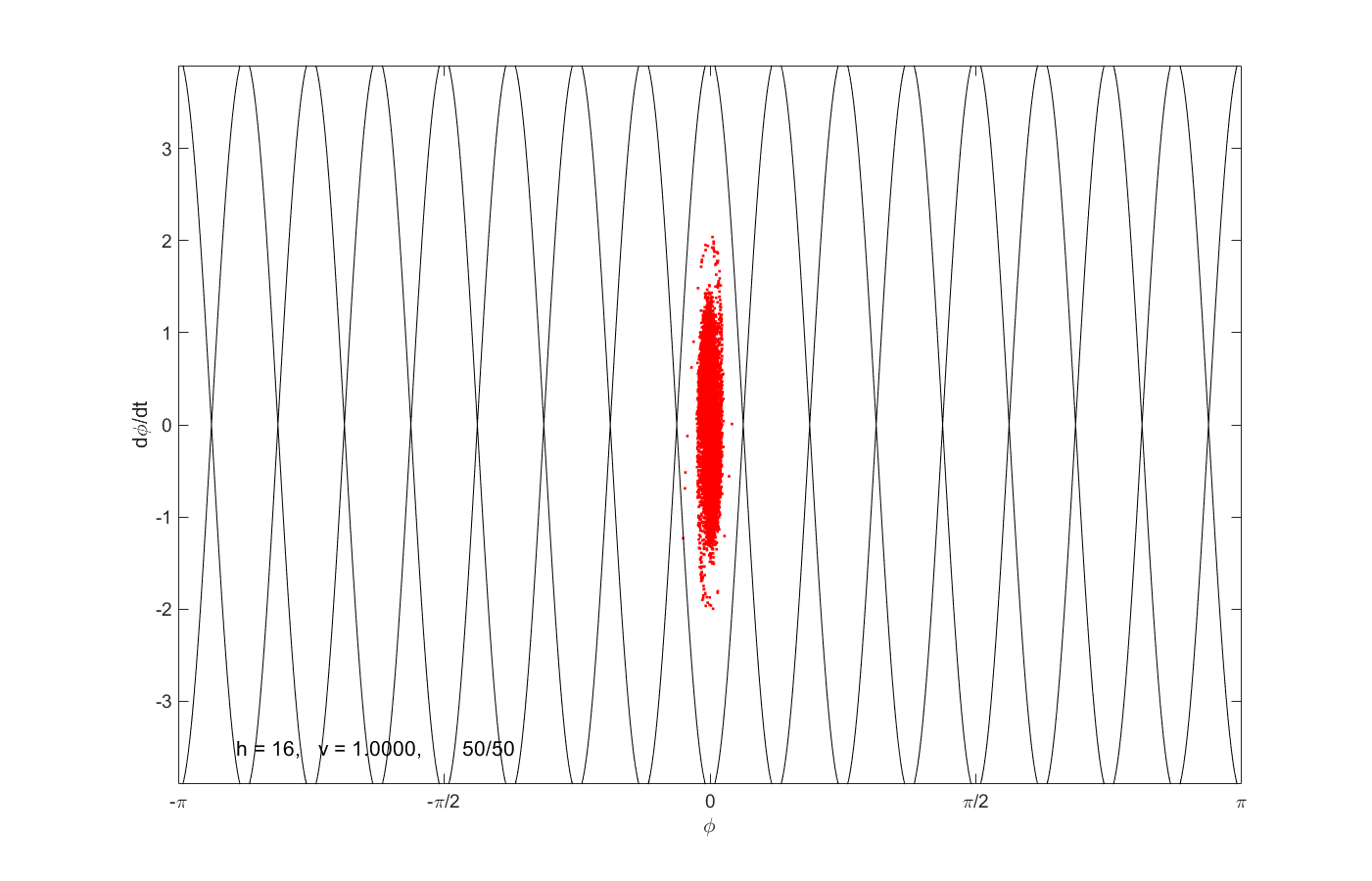}
\includegraphics[width=0.3\textwidth]{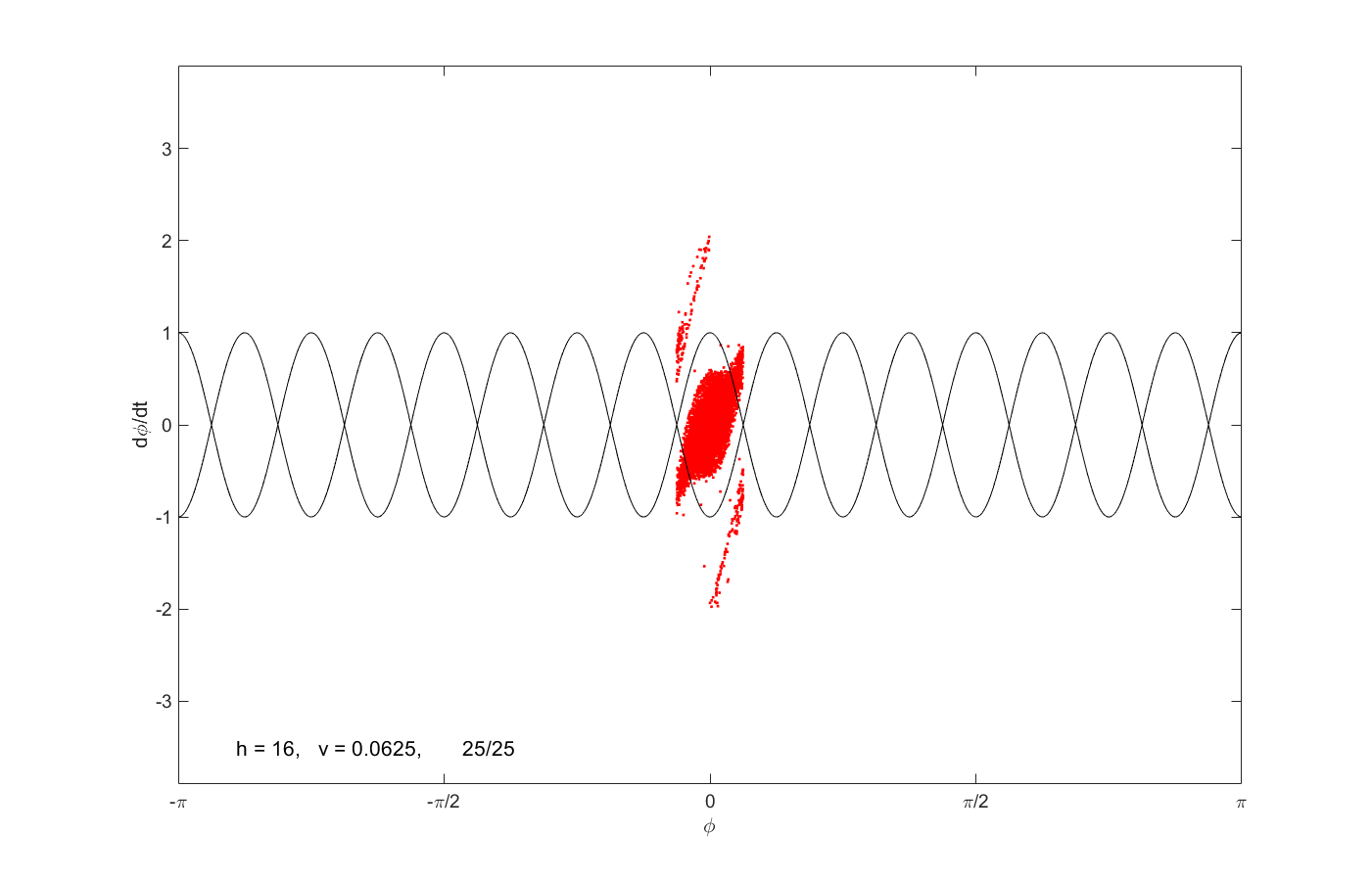}
\includegraphics[width=0.3\textwidth]{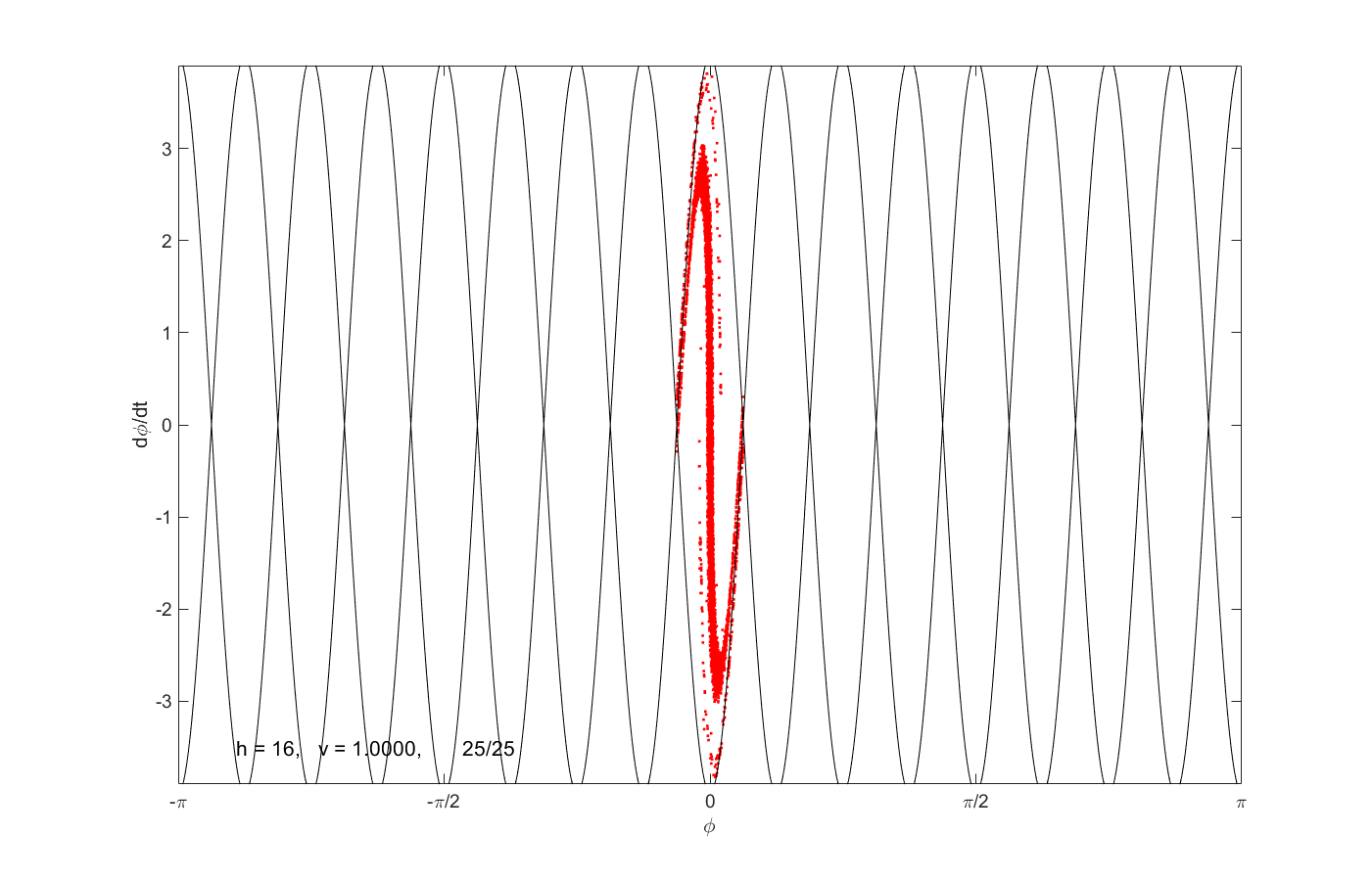}
\includegraphics[width=0.4\textwidth]{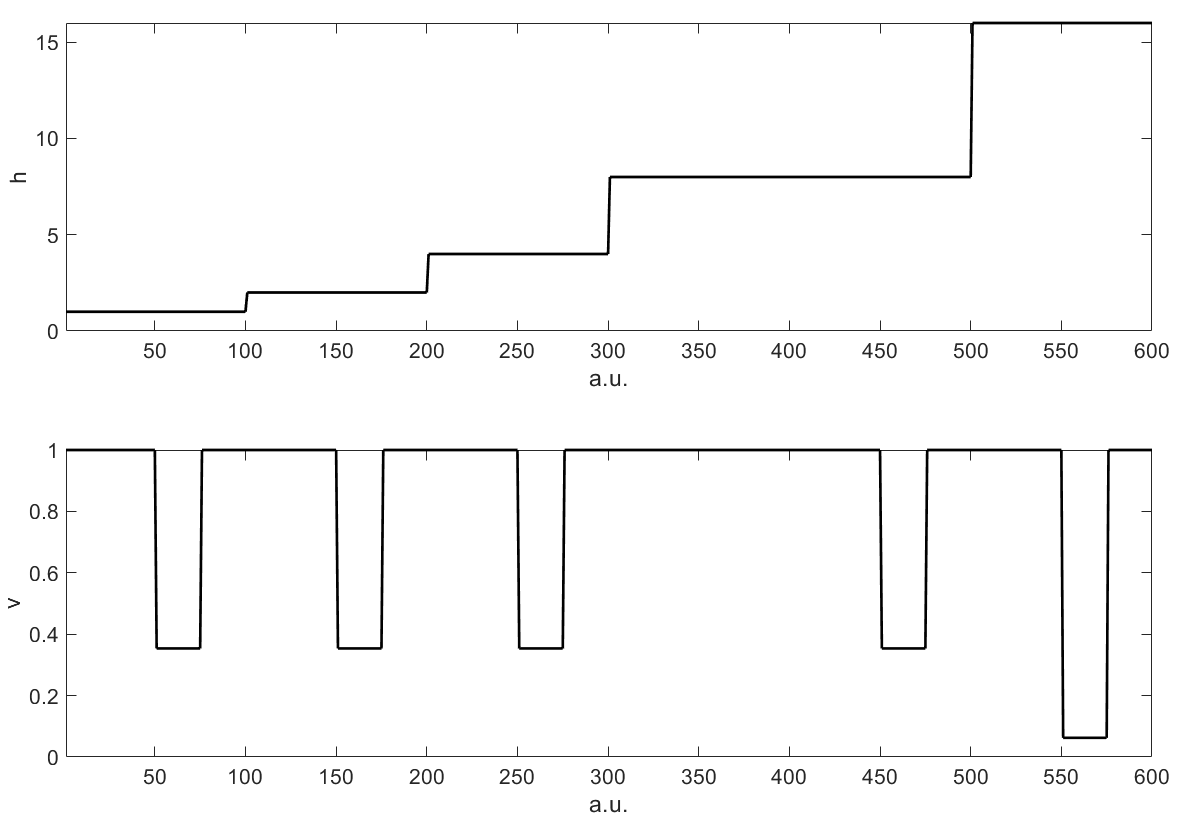}
\includegraphics[width=0.4\textwidth]{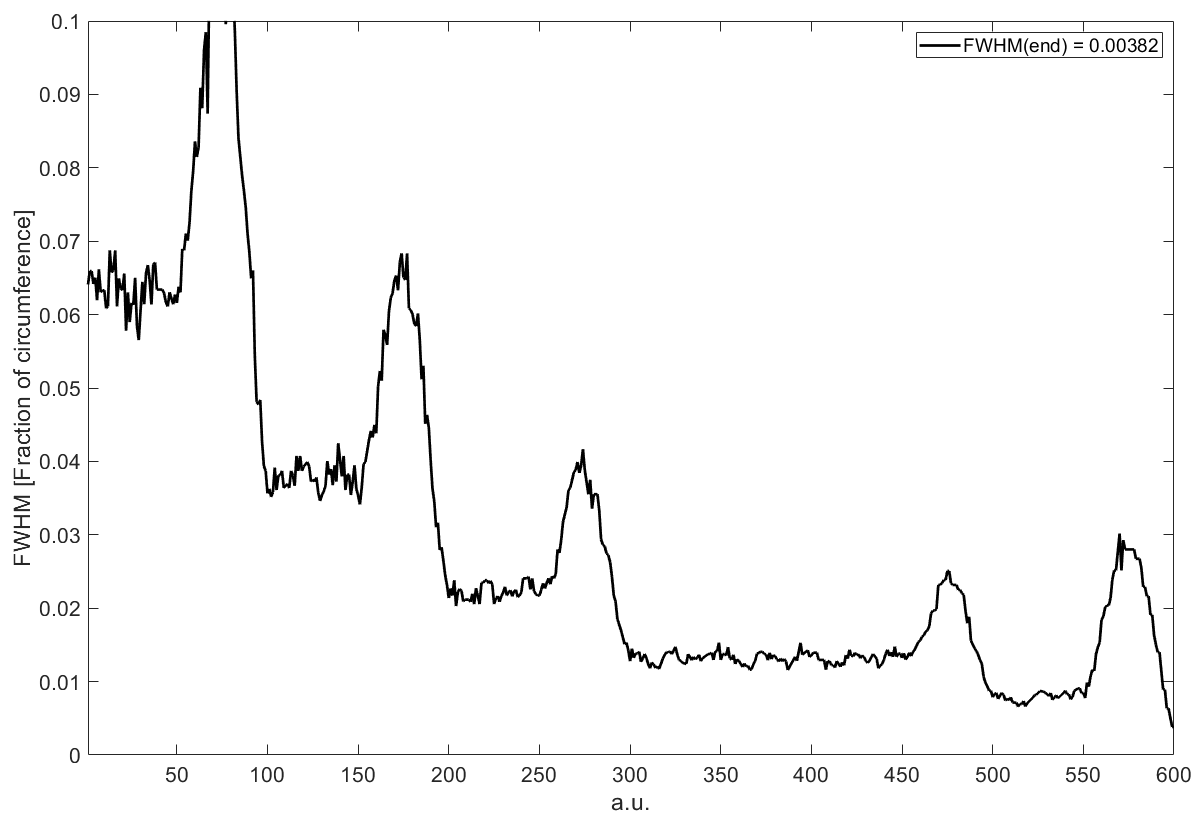}
\end{center}
\caption{\label{fig:Z}The phase-space distributions for the harmonic-doubling cascade.
  The first four rows show the munches for the sequence harmonics $h=1,2,4,8,16$, whereas
  the fifth row shows the final munch to reduce the bunch length even further. The
  bottom row shows the evolution of harmonic (upper left) and voltage (lower left) and
  the evolution of the bunch length (right).}
\end{figure}
Since very short bunches are of particular interest, both for plasma accelerators
using protons~\cite{AWAKE} and for muon colliders~\cite{MUCOLL,MUON2} we explore a
sequence of munches that pass the beam from one RF system to one at twice the
frequency and then repeat this process several times. We also assume that the
voltages of all RF systems are equal to $v=1$, only the harmonic $h$ doubles
from one stage to the next. 
\par
The result of this simulation, based on following 10000 particles, is shown in 
Figure~\ref{fig:Z}. The top-left image shows the initial bunch distribution at 
the first harmonic, which has a rms bunch length of $10^o$, which corresponds to a
FWHM of about 6.2\,\% of the circumference. The middle image in the top row shows
the distribution after
the first munch, where the voltage is dropped to $(1/2)^{3/2}\approx 0.354$ and 
on the image on the right after the final quarter oscillation, at which point it
is matched to the second-harmonic system. The second row shows the distribution
during the time it is under the control of the second-harmonic system. The left
image is taken after one half synchrotron period, the middle at the end of the
munch and the image in  the right after the quarter oscillation with the
restored voltage to prepare a matched distribution for the fourth harmonic. Note
that the height of the separatrix at the second harmonic is increased compared
to the first harmonic by the ratio of the synchrotron frequencies
$\sqrt{h_2/h_1}=\sqrt{2}$. The third row repeats the same pattern at the fourth
harmonic that we switch on as the second-harmonic system is switched off.
First we show the distribution after half a synchrotron period, then after the
munch, and the right image shows it after the quarter oscillation with
restored voltage, ready to be handed over to the eighths-harmonic system. The
fourth row repeats that pattern at the eighths-harmonic to prepare the
distribution for the sixteenth harmonic that is shown in the left image on
the fifth row. In the middle image in the fifth row, the bunch is munched a
final time with a voltage drop to $\hat v = 1/9$ such that the last quarter
oscillation reduces the bunch length a final time, ready to direct it
onto a target. At this point it has a FWHM of about 0.004 of the circumference,
which corresponds to  a few ns in a 300\,m ring. The two images on the bottom
row of Figure~\ref{fig:Z} show the evolution of the harmonic number on the
top panel of the left image and the voltage on the bottom panel. The image on
the right shows the evolution of the FWHM bunch length during the process. All
munches are clearly visible by the increased FWHM with subsequent drop to
about 60\,\% of the FWHM before the munch, a value consistent with
Equation~\ref{eq:demag}, given by $(h_1/h_2)^{3/4}=(1/2)^{3/4}\approx 0.595$.
The FWHM after the final munch is reported in the legend.
\par
We conclude that we can reduce the length of a bunch significantly, provided
that initially it is not too large. During the transfer from one harmonic
to the next, more of the stable phase space area is occupied and at some point
the non-linearity of the RF voltage will prevent a transition to an even higher
harmonic.
\section{Conclusions}
Within the framework of a linearized theory we derived equations to describe a
bunch munch consisting of two quarter-period synchrotron oscillations at two
RF voltages that transfers a longitudinally matched distribution from one RF
system to another one operating at a different harmonic and a different voltage.
Simulations taking the non-linearity of the RF systems into account show good
agreement with the linear theory; the transferred distributions are matched and
the changes of the bunch length are consistent with the linearized theory.
\par
At the same time, simulations of the non-linear system indicate limitations
of applicability. If a significant fraction of the beam distribution is
located outside the linear regime near the center of the separatrix, the
reversibility of the transfers is spoilt and the longitudinal emittance will
grow. Moreover, if the bunch area becomes comparable to the bucket size,
particles will no longer be trapped inside the same bucket and the phase space
will be diluted.
\par
We have not addressed technical aspects of implementing the munches, though
they are likely significant. Implementing the rapid changes of the RF parameter
require an advanced low-level RF control system and changes of beam loading
during the rapid changes pose additional problems. On the other hand, the
transitions are very rapid, on the order of a few synchrotron periods, such
that space-charge effects are possibly limited, though this must be
investigated in detail in the future.
%
%
\bibliographystyle{plain}

\end{document}